\RequirePackage{ifthen}

\newcommand{\typof}{1} %

\newcommand{\longv}[1]{\ifthenelse{\equal{\typof}{0}}{}{#1}}
\newcommand{\shortv}[1]{\ifthenelse{\equal{\typof}{0}}{#1}{}}
\newcommand{\longshortv}[2]{\ifthenelse{\equal{\typof}{0}}{#2}{#1}}
\newcommand{\drop}[1]{\ifthenelse{\equal{\typof}{0}}{}{}}

\documentclass[a4paper,UKenglish]{article}

\longv{
  \author{Ugo Dal Lago \and Francesco Gavazzo}
  \title{Resource Transition Systems and Full Abstraction for Linear Higher-Order Effectful Systems}
  }

\usepackage{proof}
\usepackage{amsmath}
\usepackage{amsthm}
\usepackage{tikz}
\usetikzlibrary{positioning,decorations.markings,arrows.meta,calc,fit,quotes,cd,math,arrows,backgrounds,shapes.geometric}
\usepackage{tikzit}

\usepackage{geometry}
\usepackage{bm}
\usepackage{amssymb}
\usepackage{color}
\usepackage{url}
\usepackage{stmaryrd}
\usepackage{xypic}

\newenvironment{varitemize}
{
\begin{list}{\labelitemi}
{\setlength{\itemsep}{0pt}
 \setlength{\topsep}{0pt}
 \setlength{\parsep}{0pt}
 \setlength{\partopsep}{0pt}
 \setlength{\leftmargin}{15pt}
 \setlength{\rightmargin}{0pt}
 \setlength{\itemindent}{0pt}
 \setlength{\labelsep}{5pt}
 \setlength{\labelwidth}{10pt}
}}
{
 \end{list}
}
\newcounter{numberone}
\newenvironment{varenumerate}
{
\begin{list}{\arabic{numberone}.}
{
  \usecounter{numberone}
  \setlength{\itemsep}{0pt}
  \setlength{\topsep}{0pt}
  \setlength{\parsep}{0pt}
  \setlength{\partopsep}{0pt}
  \setlength{\leftmargin}{15pt}
  \setlength{\rightmargin}{0pt}
  \setlength{\itemindent}{0pt}
  \setlength{\labelsep}{5pt}
  \setlength{\labelwidth}{15pt}
}}
{
\end{list}
} 

\newcommand{\hide}[1]{}

\newcommand{\bnf}{\;::=\;}

\newcommand{\midd}{\; \; \mbox{\Large{$\mid$}}\;\;}

\newcommand{\values}{\mathcal{V}}

\newcommand{\subst}[3]{#1[#2:=#3]}

\newcommand{\valone} {V}
\newcommand{\varone} {x}

\newcommand{\termone}{e}
\newcommand{\termtwo}{f}
\newcommand{\termthree}{g}

\newcommand{\termfour}{h}

\newcommand{\env}{\Gamma}

\newcommand{\imp}{\vdash}

\newcommand{\ctxone}{\mathcal{C}}
\newcommand{\ctxtwo}{\mathcal{D}}

\newcommand{\ectxone}{E}

\newcommand{\support}[1]{\mathsf{supp}(#1)}

\newcommand{\signature}{\Sigma}

\newcommand{\abs}[1]{\lambda #1.}

\newcommand{\ocppo}{\ensuremath{\omega\text{-}\mathsf{cppo}}}

\newcommand{\lan}{\langle}
\newcommand{\ran}{\rangle}
\newcommand{\cc}{\cdots}
\newcommand{\hh}{\hdots}

\newcommand{\redex}{r}

\newcommand{\obs}{\textnormal{\texttt{obs}}}

\newcommand{\distribution}{\mathcal{D}}

\newcommand{\inl}{\mathsf{in}_l}
\newcommand{\inr}{\mathsf{in}_r}
\newcommand{\monad}{T}
\newcommand{\unit}{\eta}

\newcommand{\defeq}{\triangleq}
\newcommand{\ctxeq}{\equiv^{\scriptscriptstyle \mathtt{ctx}}}

\newcommand{\relone}{\mathbin{R}}

\newcommand{\cpoleq}{\sqsubseteq}
\newcommand{\lub}{\bigsqcup}

\newcommand{\sem}[1]{\llbracket #1 \rrbracket}

\newcommand{\kleisli}[1]{#1^{\dagger}}
\newcommand{\return}{\mathsf{return}}

\renewcommand{\valone}{v}
\newcommand{\valtwo}{w}

\renewcommand{\termone}{e}
\renewcommand{\termtwo}{f}
\renewcommand{\termthree}{g}
\renewcommand{\termfour}{h}

\newcommand{\TO}[1]{\Rightarrow^{n}}

\newcommand{\mmap}{\multimap}

\newcommand{\comp}{\mathbin{\circ}}

\renewcommand{\monad}{T}

\newcommand{\done}{d}

\newcommand{\typeOne}{\sigma}
\newcommand{\typeTwo}{\tau}
\newcommand{\typeThree}{\rho}
\newcommand{\typeone}{\typeOne}
\newcommand{\typetwo}{\typeTwo}
\newcommand{\typethree}{\typeThree}

\newcommand{\op}{\mathbf{op}}

\newcommand{\seq}[2]{\mathbf{let}\ \varone = #1\ \mathbf{in}\ #2}

\renewcommand{\return}[1]{\mathbf{val}\ #1}
\newcommand{\val}[1]{\return{#1}}

\renewcommand{\inl}[1]{\mathbf{inl}\ #1}
\renewcommand{\inr}[1]{\mathbf{inr}\ #1}

\newcommand{\emptyenv}{\emptyset}

\newcommand{\strongkleisli}[1]{#1^{*}}

\newcommand{\divergence}{\uparrow}

\newcommand{\bang}{{!}}

\newcommand{\valimp}{\imp^{\mathsf{v}}}
\newcommand{\compimp}{\imp^{\scriptscriptstyle{\Lambda}}}

\newcommand{\fun}[2]{#1 \mapsto #2}

\newcommand{\semn}[2]{\sem{#1}_{#2}}

\newcommand{\eval}{\mathit{eval}}

\newcommand{\similarity}{\preceq}

\newcommand{\actone}{\ell}

\newcommand{\bindsymbol}{\scalebox{0.5}[1]{$>\!>=$}}
\newcommand{\bind}{\mathrel{\bindsymbol}}

\newcommand{\maybe}{M}

\renewcommand{\distribution}{D}

\newcommand{\Output}{\mathcal{O}}

\newcommand{\printsymbol}[1]{\mathbf{print}_{#1}}

\newcommand{\computations}{\Lambda}

\newcommand{\gone}{\Gamma}
\renewcommand{\done}{\Delta}
\newcommand{\envconf}{\mathcal{C}}
\newcommand{\econf}[2]{(#1;#2)}
\newcommand{\econfterm}[3]{(#1;#2;#3)}
\newcommand{\econfone}{K}
\newcommand{\econftwo}{L}

\newcommand{\sat}{\mathtt{st}}
\newcommand{\pairconf}[2]{#1:#2}
\newcommand{\cvarone}{z}
\newcommand{\envterm}[3]{#1 \mid #2 \mid #3}

\newcommand{\stuckterm}{\mathsf{s}}

\newcommand{\semf}[1]{\sem{#1}^{\scriptscriptstyle{\mathcal{F}}}}
\newcommand{\semfn}[2]{\semf{#1}_{#2}}

\renewcommand{\emptyenv}{\emptyset}
\newcommand{\semfstar}[1]{\sem{#1}^{\scriptscriptstyle{\mathcal{F}^*}}}
\newcommand{\semfstarn}[2]{\sem{#1}^{\scriptscriptstyle{\mathcal{F}^*}}_{#2}}

\newcommand{\seml}[1]{\sem{#1}^{\scriptscriptstyle{\computations}}}
\newcommand{\semln}[2]{\seml{#1}_{#2}}
\newcommand{\semlstar}[1]{\sem{#1}^{\scriptscriptstyle{\computations^*}}}
\newcommand{\syslambda}{\computations}

\newcommand{\obslambdastar}{\obs^{\scriptscriptstyle \strongkleisli{\computations}}}
\newcommand{\obsfstar}{\obs^{\scriptscriptstyle \strongkleisli{\mathcal{F}}}}
\newcommand{\obskappastar}{\obs^{\scriptscriptstyle \strongkleisli{\mathcal{K}}}}

\newcommand{\syskappa}{\mathcal{K}}
\newcommand{\kappasimilarity}{\similarity^{\scriptscriptstyle{\mathcal{K}^*}}}

\newcommand{\finsimilarity}{\similarity_{\mathtt{fin}}}

\newcommand{\finkappastarsimilarity}{\finsimilarity^{\scriptscriptstyle{\strongkleisli{\syskappa}}}}

\newcommand{\sysf}{\mathcal{F}}
\newcommand{\semk}[1]{\sem{#1}^{\scriptscriptstyle{\syskappa}}}
\newcommand{\semkn}[2]{\semn{#1}{#2}^{\scriptscriptstyle{\syskappa}}}
\newcommand{\semkstar}[1]{\sem{#1}^{\scriptscriptstyle{\strongkleisli{\syskappa}}}}
\newcommand{\semkstarn}[2]{\semn{#1}{#2}^{\scriptscriptstyle{\strongkleisli{\syskappa}}}}
\newcommand{\mstatefour}{U}
\newcommand{\kappatraceleq}{\leq_{\scriptscriptstyle{\mathtt{Tr}}}^{\scriptscriptstyle{\syskappa}}}
\newcommand{\kappatraceeq}{\simeq_{\scriptscriptstyle{\mathtt{Tr}}}^{\scriptscriptstyle{\syskappa}}}
\newcommand{\kappastartraceleq}{\leq_{\scriptscriptstyle{\mathtt{Tr}}}^{\scriptscriptstyle{\strongkleisli{\syskappa}}}}
\newcommand{\kappastartraceeq}{\simeq_{\scriptscriptstyle{\mathtt{Tr}}}^{\scriptscriptstyle{\strongkleisli{\syskappa}}}}
\newcommand{\kappastarbisimilarity}{\simeq^{\scriptscriptstyle{\strongkleisli{\syskappa}}}}
\newcommand{\lambdatraceleq}{\leq_{\scriptscriptstyle{\mathtt{Tr}}}^{\scriptscriptstyle{\syslambda}}}
\newcommand{\lambdatraceeq}{\simeq_{\scriptscriptstyle{\mathtt{Tr}}}^{\scriptscriptstyle{\syslambda}}}

\makeatletter
\newcommand{\uset}[3][0ex]{%
  \mathrel{\mathop{#3}\limits_{
    \vbox to#1{\kern-7\ex@
    \hbox{$\scriptstyle#2$}\vss}}}}
\makeatother

\newcommand{\eff}[4]{
\boxed{#1}
\raisebox{0.4em}{$\underbar{$\overset{\scriptstyle\makeset{#2}}{\quad}$ }$}
\raisebox{-0.34em}{$\overline{ \scriptstyle #3\quad  }$} #4
}

\newcommand{\effshort}[4]{
\boxed{#1}
\raisebox{0.4em}{$\underbar{$\overset{\scriptstyle\makeset{#2}}{\quad}$ }$}
\raisebox{-0.34em}{$\overline{ \scriptstyle #3\quad  }$} #4
}

\newcommand{\effj}[4]{
\boxed{#1}
\raisebox{0.4em}{$\underbar{$\overset{\scriptstyle\hspace{0.1cm}\makeset{#2}}{\quad}$ }$}
\raisebox{-0.31em}{$\overline{ \scriptstyle #3\quad  }$} #4
}

\newcommand{\efftwo}[7]{
\boxed{#1}
\raisebox{0.4em}{$\underbar{$\overset{\scriptstyle\makeset{#2}}{\quad}$ }$}
\raisebox{-0.34em}{$\overline{ \scriptstyle #3\quad  }$}
\boxed{#4}
\raisebox{0.4em}{$\underbar{$\overset{\scriptstyle\hspace{0.1cm}\makeset{#5}}{\quad}$ }$}
\raisebox{-0.34em}{$\overline{ \scriptstyle #6\quad  }$} #7
}

\newcommand{\effnull}[2]
{
\boxed{#1}
\raisebox{0.4em}{$\underbar{$\overset{\scriptstyle\phantom{n}}{\quad}$ }$}
\raisebox{-0.34em}{$\overline{ \scriptstyle \phantom{i}\quad  }$} #2
}

\newcommand{\effk}[4]{
\boxed{#1}
\raisebox{0.4em}{$\underbar{$\overset{\scriptstyle\makeset{#2}}{\quad}$ }$}
\raisebox{-0.51em}{$\overline{ \scriptstyle #3\quad  }$} #4
}

\newcommand{\effunit}[1]{
  \boxed{\eta}
\raisebox{0.4em}{$\underbar{$\overset{\scriptstyle\phantom{[1]}}{\quad}$ }$}
\raisebox{-0.34em}{$\overline{ \scriptstyle \phantom{1}\quad  }$} #1
}

\newcommand{\shorteffunit}[1]{
  \boxed{\eta}
\raisebox{0.4em}{$\underbar{$\overset{\scriptstyle\phantom{[1]}}{\quad}$ }$}
\raisebox{-0.34em}{$\overline{ \scriptstyle \phantom{1}\quad  }$} #1
}

\newcommand{\effconvergence}[1]{
 \boxed{\downarrow}
\raisebox{0.4em}{$\underbar{$\overset{\phantom{\scriptstyle[1]}}{\quad}$ }$}
\raisebox{-0.34em}{$\overline{\phantom{\scriptstyle{1}}\quad}$} #1
}
\newcommand{\boteff}{
  \boxed{\bot}
}

\newcommand{\boteffn}[1]{
  \boxed{\bot_{#1}}
}

\newcommand{\makeset}[1]{\boldsymbol{#1}}

\newcommand{\mtermone}{\xi}
\newcommand{\mtermtwo}{\varphi}

\newcommand{\typeconfone}{\alpha}
\newcommand{\typeconftwo}{\beta}
\newcommand{\typefun}{\mathtt{b}}

\newcommand{\mconfone}{\kappa}
\newcommand{\mconftwo}{\rho}
\newcommand{\traceone}{\mathsf{t}}
\newcommand{\tracetwo}{\mathsf{u}}
\newcommand{\mfstarone}{\zeta}
\newcommand{\mfstartwo}{\theta}

\newtheorem{definition}{Definition}
\newtheorem{example}{Example}
\newtheorem{theorem}{Theorem}
\newtheorem{proposition}{Proposition}
\newtheorem{remark}{Remark}
\newtheorem{lemma}{Lemma}
\newtheorem{corollary}{Corollary}

\newcommand{\evarone}{a}

\newcommand{\coseq}[2]{\mathbf{let}\ \bang{\evarone} = #1\ \mathbf{in}\ #2}
\renewcommand{\env}[2]{#1 \mid #2}
\newcommand{\lenvone}{\Delta}
\newcommand{\eenvone}{\Gamma}

\newcommand{\vect}[1]{\mathbf{#1}}

\renewcommand{\distribution}{\mathcal{D}}

\renewcommand{\varone}{x}

\renewcommand{\termone}{e}
\renewcommand{\termtwo}{f}
\renewcommand{\valone}{v}
\renewcommand{\valtwo}{w}
\renewcommand{\ctxone}{C}
\renewcommand{\ctxtwo}{D}
\newcommand{\mstateone}{\xi}
\newcommand{\mstatetwo}{\varphi}
\newcommand{\mstatethree}{Z}

\renewcommand{\maybe}{\mathcal{M}}

\renewcommand{\distribution}{\mathcal{D}}

\renewcommand{\fun}[2]{#1 \vartriangleright #2}

\newcommand{\gop}{\gamma}
\newcommand{\goptwo}{\alpha}

\renewcommand{\fun}[2]{#1 \to #2}

\newcommand{\push}{\textnormal{\texttt{push}}}

\renewcommand{\similarity}{\lesssim}

\newcommand{\bc}[1]{\mathbin{\mathtt{BC}(#1)}}
\usepackage{stackengine}
\usepackage{adjustbox}
\usepackage{mathtools}

\begin{document}

\shortv{}  

\maketitle

\begin{abstract}
We investigate program equivalence for linear higher-order
(sequential) languages endowed with primitives for computational
effects.  More specifically, we study operationally-based notions of
program equivalence for a \emph{linear} $\lambda$-calculus with
\emph{explicit copying} and \emph{algebraic effects} \emph{\`a la}
Plotkin and Power.  Such a calculus makes explicit the interaction
between copying and linearity, which are \emph{intensional} aspects of
computation, with effects, which are, instead, \emph{extensional}.  We
review some of the notions of equivalences for linear calculi proposed
in the literature and show their limitations when applied to effectful
calculi where copying is a first-class citizen.  We then introduce
\emph{resource transition systems}, namely transition systems whose
states are built over tuples of programs representing the available
resources, as an operational semantics accounting for both intensional
and extensional interactive behaviors of programs.  Our main result is
a \emph{sound and complete} characterization of contextual equivalence as
\emph{trace equivalence} defined on top of resource transition
systems.
\end{abstract}

\section{Introduction}
\label{sect:introduction}

This work aims to study operationally-based 
equivalences for 
higher-order sequential programming languages enjoying three main features, which 
we are going to explain: 
\emph{algebraic effects}, \emph{linearity}, and \emph{explicit copying}. 

\noindent\textbf{Algebraic Effects}\; Since the early days of
programming language semantics, the study of computational effects,
i.e. those aspects of computations that go beyond the pure process of
computing, has been of paramount importance. Starting with the seminal
work by Moggi \cite{Moggi/LICS/89,Moggi/Notions-Of-Computations/1991},
modelling and understanding computational effects in terms of monads
\cite{MacLane/Book/1971} has been a standard practice in the
denotational semantics of higher-order sequential languages.  More
recently, Plotkin and Power
\cite{Plotkin/algebraic-operations-and-generic-effects/2003,PlotkinPower/FOSSACS/01,DBLP:journals/entcs/PlotkinP01}
have extended the analysis of computational effects in terms of monads
to \emph{operational semantics}, introducing the theory of
\emph{algebraic effects}.  Accordingly, computational effects are
produced by effect-triggering operations whose behaviour is, in
essence, algebraic.  Examples of such operations are nondeterministic
and probabilistic choices, primitives for I/O, primitives for reading
and writing from a global store, and many others.
The operational analysis of computational effects in terms of
algebraic operations also gave new insights not only on the 
operational semantics of effectful programming languages but also
on their theories of equality, this way
leading to the development of, e.g., effectful logical relations
\cite{JohannSimpsonVoigtlander/LICS/2010,DBLP:journals/pacmpl/BiernackiPPS18},
effectful applicative and normal form/open bisimulation
\cite{DalLagoGavazzoLevy/LICS/2017,DBLP:conf/esop/LagoG19}, and
logic-based equivalences
\cite{Simpson-Niels/Modalities/2018,DBLP:conf/fossacs/MatacheS19}.

\noindent\textbf{Linearity and Copying}\;
The analysis of effectful computations in terms of monads and
algebraic effects is, in its very essence, \emph{extensional}:
ultimately, a program represents a function from inputs to monadic
outputs.  However, when reasoning about computational effects, also
\emph{intensional} aspects of programs may be relevant. In particular,
\emph{linearity}
\cite{DBLP:journals/tcs/Girard87,DBLP:conf/ifip2/Wadler90} (and its
quantitative refinements
\cite{DBLP:journals/tcs/GirardSS92,DBLP:conf/esop/GhicaS14,Brunel-et-al/ESOP/2014,DBLP:conf/lics/Atkey18,DBLP:conf/tlca/LagoH09})
has been recognised as a fundamental tool to reason about
computational effects
\cite{DBLP:journals/corr/abs-1209-4268,DBLP:journals/corr/MogelbergS14},
as witnessed by a number of programming languages, such as
\textsf{Clean} \cite{DBLP:journals/entcs/Plasmeijer95}, \textsf{Rust}
\cite{DBLP:conf/sigada/MatsakisK14}, 
\textsf{Granule} \cite{Orchard:2019:QPR:3352468.3341714}, and \textsf{Linear Haskell}
\cite{DBLP:journals/pacmpl/BernardyBNJS18}, which explicitly rely on
linearity to structure and manage effects.  Indeed, the interaction
between linearity, copying, and computational effects deeply
influences program equivalence: there are effectful programs that
cannot be discriminated without allowing the environment to copy them,
and thus program transformations which are \emph{sound} if linearity is
guaranteed, but \emph{unsound} in presence of copying.

A simple, yet instructive example of such a transformation, which we
will carefully examine in the next section, is given by distributivity
of $\lambda$-abstraction over probabilistic choice operators:
$\abs{\varone}{(\termone \oplus \termtwo)} \simeq
(\abs{\varone}{\termone}) \oplus (\abs{\varone}{\termtwo})$.  This
transformation is well-known to be unsound for `classical'
call-by-value probabilistic languages
\cite{CrubilleDalLago/ESOP/2014}. However, it is sound
if the programs involved cannot be 
copied \cite{DBLP:journals/tcs/DengZ15,DBLP:conf/tase/DengF17}.  
What, instead, we expect
to be unsound is the transformation $\bang{(\termone \oplus \termtwo)}
\simeq \bang{\termone} \oplus \bang{\termtwo}$, where the operator
$\bang$ (bang) is the usual linear logic exponential modality making
terms under its scope copyable and erasable.  It is thus natural to
ask if, and to what extent, the aforementioned notions of effectful
program equivalence can be extended to \emph{linear} languages with
\emph{explicit copying}.

\noindent\textbf{Our Contribution}\;
In this paper we introduce \emph{resource transition systems} as an
intensional, resource-sensitive operational semantics for linear
languages with algebraic operations and explicit copying.  Resource
transition systems combine standard \emph{extensional} properties of
effectful computations with linearity and copying, whose nature is,
instead, \emph{intensional}.  We model the former using monads---as one
does for ordinary effectful semantics---and the latter by shifting from
program-based transition systems to \emph{tuple-based} transition systems, as
one does in environmental bisimulation
\cite{Sangiorgi/Environmental/2011,DBLP:conf/concur/MadiotPS14}.
Indeed, a resource transition system can be thought of as an ordinary
transition system whose states are built over tuples of copyable
programs and linear values representing the available resources
produced by a program while interacting with the external
environment. Another possible way to look at resource transition
systems is as an interactive semantics defined on top of the so-called storage
model \cite{DBLP:journals/tcs/TurnerW99}.
We then define and study trace equivalence on resource transition systems. 
Our main result states trace equivalence is \emph{sound} and
\emph{complete} for contextual equivalence. To the best of the
authors' knowledge, this is the first full abstraction result for a
linear $\lambda$-calculus with arbitrary algebraic effects and
explicit copying.

\noindent\emph{Outline}\; 
This paper is structured as follows.  After an informal introduction
to program equivalence for effectful linear languages
(Section~\ref{sect:informal-example}), Section~\ref{sect:monads} 
recalls some background
notions on monads and algebraic operations.
Section~\ref{sect:environmental-equivalences} introduces our vehicle
calculus and its operational semantics. 
Resource-sensitive resource transition systems and their associated notions of 
equivalence are given in Section~\ref{section:resource-sensitive-transition-systems}.

\section{Effects, Linearity, and Program Equivalence}
\label{sect:informal-example}

In this section, we give a gentle introduction to program
equivalence in presence of linearity, explicit copying, and
effects. In this work, we are concerned with 
 \emph{operationally-based} 
equivalences, example of those being contextual and CIU equivalence 
\cite{Morris/PhDThesis,MasonTalcott/1991}, logical relations \cite{Reynolds/Logical-relations/1983,Plotkin-Lambda-definability-logical-relations,Sieber} and, 
bisimulation-based equivalences \cite{Abramsky/RTFP/1990,Lassen/BismulationUntypedLambdaCalculusBohmTrees,Lassen/EagerNormalFormBisimulation/2005,Sangiorgi/Environmental/2011}. Moreover, among operationally-based equivalences, we 
seek for lightweight ones, by which we mean equivalences which are
as easy to use as possible (otherwise, contextual equivalence would be
enough).  Accordingly, we do not consider equivalences in the spirit
of logical relations---which usually require heavy techniques such as
biorthogonality \cite{DBLP:journals/mscs/Pitts00} and step-indexing
\cite{AppelMcAllester/TOPLAS/2001} when applied to calculi in which
recursion is present, either at the level of types or at the level of
terms.  Instead, we focus on \emph{first-order} 
equivalences \cite{DBLP:conf/concur/MadiotPS14}, viz. notions of trace
equivalence and bisimilarity.

Our running examples in this paper are the already mentioned 
distributivity of (lambda) abstraction  
and bang over 
(fair) probabilistic choice in probabilistic call-by-value $\lambda$-calculi 
\cite{DalLagoSangiorgiAlberti/POPL/2014,CrubilleDalLago/ESOP/2017,DBLP:journals/tcs/DengZ15}:
\vspace{-0.3cm}
\begin{center}
\begin{minipage}{.5\textwidth}
  \begin{equation}
    \tag{$\lambda$-dist} \label{lambda-dist}
    \abs{\varone}{(\termone \oplus \termtwo)} 
    \simeq (\abs{\varone}{\termone}) \oplus (\abs{\varone}{\termtwo})
  \end{equation}
\end{minipage}
$\qquad$
\begin{minipage}{.4\textwidth}
  \begin{equation}
    \tag{$!$-dist} \label{bang-dist}
    \bang{(\termone \oplus \termtwo)} \simeq 
    \bang{\termone} \oplus \bang{\termtwo}
  \end{equation}
\end{minipage}
\end{center}

It is well-known \cite{CrubilleDalLago/ESOP/2014} 
that in call-by-value probabilistic languages, 
lambda abstraction does not distribute over probabilistic choice. 
In a linear setting, however, we see that any 
resource-sensitive notion of program equivalence $\simeq$ should actually 
validate the equivalence 
\eqref{lambda-dist}
but not
\eqref{bang-dist}.
Why? 
Let us look at the transition systems describing 
the (interactive) behaviour 
(Figure~\ref{fig:lambda-dist}) of the programs involved in \eqref{lambda-dist}.
\begin{figure*}[htbp]
\hrule 
\[
\xymatrix@R=0.8cm@C=-0.5pc{
          & \abs{\varone}{(\termone \oplus \termtwo)}
             \ar[d]_{\eval} & \\
          & \abs{\varone}{(\termone \oplus \termtwo)} \ar[d]^{@\valone}
          \\
           & \ar@{.}[ld]_{0.5} 
            \ar@{.}[rd]^{0.5} 
          & 
          \\
\subst{\termone}{\varone}{\valone}      &   & \subst{\termtwo}{\varone}{\valone}
}
\qquad \qquad
\xymatrix@R=0.8cm@C=-2pc{
          & (\abs{\varone}{\termone}) \oplus 
            (\abs{\varone}{\termtwo}) 
            \ar[d]^{\eval} 
            &
        \\
          & \ar@{.}[ld]_{0.5} \ar@{.}[rd]^{0.5} & 
          \\
\abs{\varone}{\termone} \ar[d]_{@\valone} &   & \abs{\varone}{\termtwo} \ar[d]^{@\valone} 
\\
\subst{\termone}{\varone}{\valone}      &   & \subst{\termtwo}{\varone}{\valone}
}
\]
\hrule
\caption{Interactive behaviour of $\abs{\varone}{(\termone \oplus \termtwo)}$
and 
$(\abs{\varone}{\termone}) \oplus (\abs{\varone}{\termtwo})$}
\label{fig:lambda-dist}
\end{figure*}
One way to understand the failure of the equivalence 
\eqref{lambda-dist}
in \emph{classical} (i.e. resource-agnostic) languages
is that several notions of probabilistic program equivalence 
(such as probabilistic contextual equivalence \cite{DalLagoSangiorgiAlberti/POPL/2014}, 
applicative 
bisimilarity \cite{CrubilleDalLago/ESOP/2014,DalLagoSangiorgiAlberti/POPL/2014}, and 
logical relations \cite{BizjakBirkedal/FOSSACS/2015}) 
are sensitive to branching. 
However, sensitivity to
branching does not quite feel like the crux of the failure of 
of distributivity of abstraction over choice in classical languages. 
In fact, what we see is that 
$\abs{\varone}{(\termone \oplus \termtwo)}$
waits for an input, and then resolves the probabilistic choice. Dually, 
$(\abs{\varone}{\termone}) \oplus (\abs{\varone}{\termtwo})$
first resolves the choice, and then waits for an input. 
As a consequence, if we evaluate these programs, 
$\abs{\varone}{(\termone \oplus \termtwo)}$ essentially does nothing, 
whereas 
$(\abs{\varone}{\termone}) \oplus (\abs{\varone}{\termtwo})$ 
probabilistically chooses if continuing with either 
$\abs{\varone}{\termone}$ or $\abs{\varone}{\termtwo}$. At this point, 
there is a crucial difference between the programs obtained: 
$\abs{\varone}{(\termone \oplus \termtwo)}$ still has to resolve the 
probabilistic choice. If we were allowed to pass it an argument, say $\valone$,
\emph{twice}---
this way resolving the choice twice---then 
we could observe a (probabilistic) behaviour different from both the one of 
$\abs{\varone}{\termone}$ and of $\abs{\varone}{\termtwo}$.
Indeed, assuming $\subst{\termtwo}{\varone}{\valone}$ to diverge and 
$\subst{\termone}{\varone}{\valone}$ to converge (with probability $1$), 
then, we would converge (to $\subst{\termone}{\varone}{\valone}$) 
with probability $0.25$, in the former case, 
and with probability $0.5$, in the latter case. 
To observe such a behaviour, however, 
it is crucial to \emph{copy} $\abs{\varone}{(\termone \oplus \termtwo)}$. 
Otherwise, we could only interact with it by passing it an argument only \emph{once}, 
this way validating \eqref{lambda-dist}. 

Summing up, to invalidate \eqref{lambda-dist} one has to be able to 
\emph{copy} the results of the evaluation of the programs involved. 
This observation suggests that the deep reason why \eqref{lambda-dist} fails 
relies on the copying capabilities of the calculus \cite{SangiorgiVignudelli/2016}. 
If the calculus at hand is linear (and thus offers no copying capability), 
we should then expect  \eqref{lambda-dist} to hold, while  
$\bang{\abs{\varone}{(\termone \oplus \termtwo)}} \simeq \bang{(\abs{\varone}{\termone})} \oplus \bang{(\abs{\varone}{\termtwo})}$ (and thus ultimately \eqref{bang-dist}) 
to fail.
This agrees with a recent result by Deng and Zhang 
\cite{DBLP:journals/tcs/DengZ15,DBLP:conf/tase/DengF17}, 
who observed that if a calculus does not 
have copying capabilities, then contextual equivalence 
(which is \emph{a fortiori} linear) validates
\eqref{lambda-dist}. More generally, 
Deng and Zhang showed that 
\emph{linear contextual equivalence}, i.e. 
contextual equivalence where contexts test their arguments 
linearly (viz. exactly once), coincides with
\emph{linear trace equivalence} in probabilistic languages.

But what about \eqref{bang-dist}? Unfortunately, 
linear trace equivalence has been designed 
for linear languages \emph{without} copying, only. Moreover, 
straightforward extensions of linear 
trace equivalence to languages with copying
would actually validate \eqref{bang-dist}, trace equivalence being 
insensitive to branching.  
The situation does not change much if one looks at different forms 
of equivalence, such as Bierman's applicative 
bisimilarity~\cite{DBLP:journals/jfp/Bierman00}. Such equivalences usually 
invalidates \eqref{bang-dist}, but they all invalidate \eqref{lambda-dist} too.
We interpret all of this as a symptom of the lack of intensional structure 
in the aforementioned notions of equivalence. Ultimately, this can be 
traced back to the very operational semantics of the calculus, which 
is meant to be an abstract description of the input-output behaviour of programs, 
but gives no insight into their \emph{intensional} structure, i.e. linearity 
and copying in our case \cite{DBLP:journals/tcs/TurnerW99}.

We propose to overcome this deficiency by giving calculi a 
\emph{resource-sensitive} operational semantics on top of which 
notions of program equivalence accounting for both intensional and 
extensional aspects of programs can be naturally defined. 
We do so by shifting from program-based transition systems 
to transition systems whose states are tuples $\econf{\gone}{\done}$, where $\gone$ is a sequence 
of \emph{non-linear} (hence copyable) programs and 
$\done$ is a sequence of \emph{linear} values, as states. 
Accordingly, fixed a tuple $\econf{\gone}{\done}$ and a program $\termone$, 
we evaluate $\termone$, say obtaining a value $\valone$, and 
add $\valone$ to the linear environment 
$\done$, this way describing the \emph{extensional} behaviour of the program. 
There are two \emph{intensional} actions we can make on tuples.
If $\done$ contains a value of the form $\bang{\termone}$, then
we can remove $\bang{\termone}$ from 
$\done$ and add $\termone$ to $\gone$. Dually, once we have a program 
$\termone$ in $\gone$, we can decide to evaluate it---and thus to possibly produce 
a new linear value---\emph{without} removing it from $\gone$, this way reflecting 
its non-linear nature.
Finally, we can interact with a value $\abs{\varone}{\termtwo}$ by passing 
it an argument built using programs in $\gone$ and values in $\done$. 
As the latter are linear, we will then remove them from $\done$. 

We conclude this section by remarking that although here we have focused 
on probabilistic languages, a similar analysis can be made for 
languages exhibiting different kinds of effects, such as input-output behaviours 
as well as combinations of effects (e.g. probabilistic nondeterminism and 
global stores).

\section{Preliminaries: Monads and Algebraic Effects}
\label{sect:monads}
Starting with the seminal work by Moggi 
\cite{Moggi/LICS/89,Moggi/Notions-Of-Computations/1991}, 
\emph{monads} have become a standard 
formalism to model and study computational effects in higher-order sequential 
languages. Instead of working with monads, we opt for the 
equivalent notion of a \emph{Kleisli triple} \cite{MacLane/Book/1971}.

\begin{definition}
\label{def:monad}
A \emph{Kleisli triple} is triple $(\monad, \unit, \bind)$ 
consisting of a map associating to any set $X$ a set $\monad(X)$, 
a set-indexed family of functions $\unit_X: X \to \monad(X)$, and 
a map $\bind$, called \emph{bind}, associating 
to each function $f: X \to \monad(Y)$ a function $\bindsymbol{f}: \monad(X) \to \monad(Y)$. 
Additionally, these data must obey the following laws, for 
$f$ and $g$ functions with appropriate (co)domains:
\begin{align*}
\bindsymbol{\unit} &= id;
&
\bindsymbol{f} \comp \unit &= f;
&
\bindsymbol{g} \comp \bindsymbol{f}
&= \bindsymbol{(\bindsymbol{g} \comp f)}.
\end{align*}
Following standard practice, we write $m \bind f$ for $\bindsymbol f(m)$.
\end{definition}

The computational interpretation behind Kleisli triples is the following: if 
$A$ is a set (or type) of values, then $\monad(A)$ represent the set of computations 
returning values in $A$. Accordingly, for each set $A$ there is 
a function $\unit_A: A \to \monad(A)$ that 
regards a value $a \in A$ as a trivial computation returning $a$ (and producing no 
effect). The map $\unit$ corresponds to the programming constructor 
$\mathbf{return}$.
Similarly, $\mu \bind f$ is the \emph{sequential composition} of a computation 
$\mu \in \monad(A)$ with a function $f: A \to \monad(B)$, 
and corresponds to the sequencing constructor $\seq{-}{-}$. 
Following this interpretation, we can read the identities in
Definition~\ref{def:monad} as stipulating that $\unit$ indeed produces no effect, 
and that sequencing is associative. 

Monads alone are not enough to produce actual effectful computations, 
as they only provide primitives to produce trivial effects (via the map $\unit$) 
and to (sequentially) compose them (via binding). For this reason, we  
endow
monads $\monad$ with (finitary) operations, i.e. with set-indexed families of 
functions $\op_X: \monad(X)^n \to \monad(X)$, where $n \in \mathbb{N}$ is the arity 
of the operation $\op$.

\begin{example}
\label{ex:monads}
Here are examples of monads modeling some of the computational effects discussed 
in Section~\ref{sect:introduction}. 
Further examples, such as 
global stores and exceptions
can be found in, e.g., 
\cite{Moggi/LICS/89,DBLP:conf/afp/Wadler95}.
\begin{varenumerate}
  \item
We model possibly divergent computations using the 
maybe monad $\maybe(X) \defeq X + \{\divergence\}$. 
An element in $\maybe(A)$ is either an element $a \in A$ 
(meaning that we have a terminating computation returning $a$), or 
the element $\divergence$ (meaning that the computation diverges).
Given $a \in A$, the map $\unit_A$ simply (left) injects $a$ in $\maybe(A)$, whereas 
$\bindsymbol{f}$ sends a terminating computation returning $a$ to $f(a)$, 
and divergence to divergence:
\begin{align*}
\inr(a) \bind f &\defeq f(a);
&
\inr(\divergence) \bind f &\defeq \inr(\divergence).
\end{align*} 
As non-termination is an intrinsic feature of complete programming languages, 
we do not consider 
explicit operations to produce divergence. 
 \item We model probabilistic computations using the (discrete) 
    subdistribution monad $\distribution$.
    Recall that a discrete subdistribution over a \emph{countable} set $X$ is a function 
    $\mu: X \to [0,1]$ such that $\sum_{x} \mu(x) \leq 1$. 
    An element element $\mu \in \distribution(A)$ gives 
    for any $a \in A$ the probability $\mu(a)$ of returning $a$. Notice that working with 
    subdistribution we can easily model divergent computations \cite{DalLagoZorzi/TIA/2012}. 
    Given $a \in A$, $\unit_A(a)$ is the Dirac distribution on $a$ 
    (mapping $a$ to $1$ and all other 
    elements to $0$), whereas 
    for $\mu \in \distribution(A)$ and $f: A \to \distribution(B)$ we define 
    $(\mu \bind f)(b) \defeq \sum_a \mu(a) \cdot f(a)(b)$. 
    Finally, we generate probabilistic computations using a binary fair probabilistic 
    choice operation $\oplus$ thus defined: $(\mu \oplus \nu)(x) \defeq 0.5 \cdot \mu(x) + 
    0.5\cdot \nu(x)$.
\item
We model computations with output using 
    the output monad 
    $\Output(X) \defeq O^{\infty} \times (X + \{\divergence\})$, 
    where $O^{\infty}$ is the set of finite and infinite strings 
    over a fixed output alphabet $O$ and $\divergence$ is a special symbol denoting 
    divergence. 
    An element of $\Output(A)$ is either a pair $(o,\inl a)$, with $a \in A$, 
    or a pair $(o, \inr \divergence)$. The former case denotes convergence to $a$ 
    outputting $o$ (in which case $o$ is a \emph{finite} string), 
    whereas the former denotes divergence outputting $o$ (in which case 
    $o$ can be either finite or \emph{infinite}).
    Given $a \in A$, the pair $(\varepsilon, \inr a)$ 
    represents the trivial computation that returns $a$ and outputs nothing 
    ($\varepsilon$ denotes the empty string). Further, 
    sequential composition of computations is defined using string concatenation 
    as follows, where $f(a) = (o',x)$: 
    \begin{align*}
    (o,\inr \divergence) \bind f &\defeq (o,\inr \divergence);
    &
    (o,\inl a) \bind f &\defeq (o o', \nu).
    \end{align*}
    Finally, we produce outputs using (a $O$-indexed family of) 
    unary operations 
    $\printsymbol{c}$ mapping $(o,x)$ to $(c o, x)$.
 \item We model computations with input using the input monad 
  $\mathcal{I}(X) = \mu \alpha. (X + \{\divergence\}) + \alpha^I$, 
  where $I$ is an input alphabet 
  (for simplicity, we take $I = \{\mathit{true}, \mathit{false}\}$). An element in 
  $\mathcal{I}(A)$ is a binary tree whose leaves are labeled either by 
  elements in $A$ or by the divergent symbol $\divergence$. The trivial computation 
  returning $a$ is the single leaf labeled by $a$, whereas given a tree $t \in 
  \mathcal{I}(A)$ and a map $f: A \to \mathcal{I}(B)$, the tree $t \bind f$ is 
  defined by replacing the leaves of $t$ labeled by elements $a \in A$ with 
  $f(a)$. Finally, we consider a binary input operation whereby
  $\mathbf{read}(t_{\mathit{true}}, t_{\mathit{false}})$ is the 
  tree whose left child is $t_{\mathit{true}}$ and whose right child is 
  $t_{\mathit{false}}$.  
\end{varenumerate}
\end{example}

\subsection{Algebraic Effects} 
Following Example~\ref{ex:monads}, let us consider 
a probabilistic program $\termone \defeq E[\termone_1 \oplus \termone_2]$, where $E$ is an evaluation context.
The operational behavior of $\termone$
is to fairly choose a $\termone_i \in \{\termone_1, \termone_2\}$, and then
execute $E[\termone_i]$. That is, 
$E[\termone_1 \oplus \termone_2]$ evaluates to $E[\termone_1]$ (resp. $E[\termone_2]$) 
with probability $0.5$.
But that is exactly the behavior of 
$E[\termone_1] \oplus E[\termone_2]$, so that we have the program equivalence 
$E[\termone_1 \oplus \termone_2] \equiv E[\termone_1] \oplus E[\termone_2]$. 
It does not take much to realize that a similar equivalence holds for all 
operations in Example~\ref{ex:monads}.
Semantically, operations justifying these equivalences are known as 
\emph{algebraic operations} 
\cite{DBLP:journals/entcs/PlotkinP01,PlotkinPower/FOSSACS/02}.

\begin{definition}
\label{def:algebraic-operation}
An $n$-ary (set-indexed family of) operation(s)
$\op_X: \monad(X)^n \to \monad(X)$ is an \emph{algebraic operation} on $\monad$, 
if for all $X,Y$, $f: X \to \monad(Y)$, and $\mu_1, \hh, \mu_n \in \monad(X)$,  
we have:
\begin{align*}
(\op_X(\mu_1, \hh, \mu_n)) \bind f = \op_Y(\mu_1 \bind f, \hh, \mu_n \bind f).
\end{align*}
\end{definition}

Using algebraic operations we can model a large class of effects, including 
those of Example~\ref{ex:monads}, pure nondeterminism (using the powerset 
monad and set-theoretic union as binary nondeterminism choice), imperative 
computations (using the global states monad and operations for reading and 
updating stores), as well as combinations thereof \cite{DBLP:journals/tcs/HylandPP06}.

\subsection{Continuity}

Another feature shared by all monads in Example~\ref{ex:monads} is that 
they all endow sets $\monad(X)$ with an $\omega$-complete pointed partial order 
($\omega$-cppo, for short) structure making $\bind$ strict, 
monotone, and continuous in both arguments, and 
algebraic operations monotone and continuous in all arguments. This property has been formalized 
in \cite{DalLagoGavazzoLevy/LICS/2017} as $\signature$-\emph{continuity}.

\begin{definition}
Let $\monad$ be a monad and $\signature$ be a set of algebraic operations 
on $\monad$. We say that $\monad$ is $\signature$-\emph{continuous} if for any
set $X$, $\monad(X)$ carries an $\omega$-cppo structure such that
$\bindsymbol$ is strict, monotone, and continuous in both arguments, 
and (algebraic) operations in $\signature$ are monotone and continuous in 
all arguments.
\end{definition}

\begin{example}
\begin{varenumerate}
 \item The maybe monad is $\emptyset$-continuous, with $\maybe(X)$ 
   endowed with the flat order. 
\item The subdistribution monad is $\{\oplus\}$-continuous, 
  with subdistributions ordered pointwise (i.e. $\mu \leq \nu$ if and only if
  $\mu(x) \leq \nu(x)$, for any $x \in X$). 
\item Let $\signature \defeq \{\printsymbol{c} \mid c \in O\}$. 
  Then, the output monad is $\signature$-continuous, with $\Output(A)$ 
  endowed with the order:
  $
  (o,x)\cpoleq (o',x')$ if and only either 
  $x = \inr \divergence \text{ and } o \cpoleq o'$
  or $x = \inl a = x' \text{ and } o=o'$.
\item The input monad is $\{\mathbf{read}\}$-continuous with respect 
  to the standard tree ordering. 
\end{varenumerate}
\end{example}

\longv{
\section{Generic Effects}
\label{sect:diagrammatic-notation}

Representing effectful computations as monadic objects has the major
advantage of providing semantical information on the effects
performed.  However, it also has the drawback of lacking a clear
distinction between the \emph{effects} produced by a computation and
the possible \emph{results} returned. Nonetheless, 
since effects can only be produced by
(algebraic) operations, we can always decouple the effects produced 
during a computation from its possible results. This is done 
relying on the notion of a \emph{generic effect} \cite{Plotkin/algebraic-operations-and-generic-effects/2003}, which we introduce 
by means of an example. 

\begin{example}
\label{example:generic-effects-probabilistic-computations}
Recall that we model probabilistic computations using the subdistribution monad 
$\distribution$. When working with (discrete) subdistribution, it is oftentimes 
convenient to employ syntactic representations of such (sub)distributions, known as
\emph{formal sums}. 
A formal sum (over a set $X$) is an expression of the form 
$\sum_{i \in I} p_i;x_i$, where $I$ is a countable set,
$p_i \in [0,1]$, $x_i \in X$, and $\sum_i p_i \leq 1$. 
The notation $\sum_{i \in I} p_i;x_i$ is meant to recall the semantic 
counterpart of formal sums, namely subdistributions. However, we should keep 
in mind that formal sums are purely syntactical expressions. For instance, 
$\frac{1}{2}; x + \frac{1}{2};x$ and $1;x$ are two distinct formal sums, 
although they both denote the Dirac distribution on $x$.
More generally, there is an interpretation function $\mathcal{I}$ 
mapping each formal sum 
$\sum_{i \in I} p_i;x_i$ to a subdistribution 
$\mu$ on $X$ defined as $\mu(x) \defeq \sum_{x_i = x} p_i$. 
Examining a bit more carefully a formal sum 
$\sum_{i \in I} p_i;x_i$, we see that the latter consists of an $I$-indexed sequence 
$\vect{p} = \lan p_i \ran_{i \in I}$ of elements in $[0,1]$ together with an $I$-indexed sequence 
$\vect{x} = \lan x_i \ran_{i \in I}$ of elements in $X$. 
Therefore, a formal sum is just a pair of sequences 
$(\vect{p}, \vect{x}) \in [0,1]^I \times X^I$ 
such that $\sum_i p_i \leq 1$. But the latter requirement means precisely that 
$\vect{p}$ is actually a subdistribution on $I$ (the one mapping 
$i$ to $p_i$). Therefore, we see that formal sums are just elements in 
$\distribution(I) \times X^I$.
Putting these observations together, we see that for any $\mu \in \distribution(X)$, 
there exists a countable set $I$ and an element $\phi \in \distribution(I) \times X^I$ 
such that $\mathcal{I}(\phi) = \mu$. As a consequence, stipulating two formal sums 
$\phi_1, \phi_2 \in \distribution(I) \times X^I$ to be equal 
(notation $\phi_1 =_{\mathcal{I}} \phi_2$)
 if $\mathcal{I}(\phi_1) = \mathcal{I}(\phi_2)$, then we see that 
 $\distribution(X)$ is 
 isomorphic to the quotient set 
 $(\bigcup_{I} \distribution(I) \times X^I)/=_{\mathcal{I}}$, where 
$I$ ranges over countable sets.
\end{example}

Can we generalize Example~\ref{example:generic-effects-probabilistic-computations} 
to arbitrary $\signature$-continuous monads? If monads are countable, 
\cite{DBLP:journals/entcs/Power06a,10.1007/978-3-642-99902-4_3,DBLP:journals/entcs/HylandP07}
(as they are all the monads considered in this work), the answer to this question is in the affirmative. 
First, let us observe that since the set $I$ in 
Example~\ref{example:generic-effects-probabilistic-computations} is countable, we can 
replace it with an enumeration of its elements. That is, we replace $I$ with 
sets $\makeset{n}$, where $n \in \mathbb{N}^{\infty} \defeq \mathbb{N} \cup \{\omega\}$ 
and $\makeset{n} \defeq \{1, \hh, n\}$ if $n \neq \omega$, and $\makeset{n} \defeq 
\mathbb{N}_{\geq 1}$, if $n = \omega$. 

\begin{theorem}[\cite{DBLP:journals/entcs/Power06a,10.1007/978-3-642-99902-4_3,DBLP:journals/entcs/HylandP07}]
\label{thm:representation}
Let $\monad$ be countable monad. Then, 
for any countable set $X$, all elements in $\monad(X)$ can be (non-uniquely) 
presented as elements in 
$$
\bigcup_{n \in \mathbb{N}^{\infty}} \monad(\makeset{n}) \times X^n
$$
Moreover, the map $\mathcal{I}: \bigcup_{n \in \mathbb{N}^{\infty}} \monad(\makeset{n}) 
\times X^n \to \monad(X)$ mapping $(\gop, \vect{x})$ to $\kleisli{\vect{x}}(\gop)$ 
is surjection whose kernel $=_{\mathcal{I}}$ gives an isomorphism
$\monad(X) \cong \bigcup_{n \in \mathbb{N}^{\infty}} \monad(\makeset{n})/{=_{\mathcal{I}}}$.
\end{theorem}

Given a pair 
$(\gop, \vect{x}) \in 
\bigcup_{n \in \mathbb{N}^{\infty}} \monad(\makeset{n})/{=_{\mathcal{I}}}$, 
we think about $\gop$ as the effect produced during a computation, and about 
$\vect{x}$ as the possible values returned.  
Elements in $\monad (\makeset{n})$ are called 
\emph{generic effects} \cite{Plotkin/algebraic-operations-and-generic-effects/2003}, 
whereas we refer to the set $\{x_i\}_{i \in \makeset{n}}$ associated to 
$\vect{x}$ as the \emph{support} of 
$\mtermone$. 

By Theorem~\ref{thm:representation}, we can represent any monadic element 
as an equivalence class of a pair $(\gop, \vect{x})$. Working with 
such pairs\footnote{For simplicity, we work with pairs $(\gop, \vect{x})$
rather than with their equivalence classes: it is a straightforward exercise 
to check that all definitions we give relying on such pairs do not actually 
depend on the specific choice of the pair, so that they extend to 
elements in $\bigcup_{n \in \mathbb{N}^{\infty}} \monad(\makeset{n})/{=_{\mathcal{I}}}$.}
allows us to simplify proofs. Moreover, elements in $\monad(\makeset{n})$ 
form an \emph{operand} \cite{leinster_2004, place={Cambridge}}. In particular, 
they come with a notion of composition that mapping
all generic effects
$\gop \in \monad(\makeset{n})$, $\goptwo_1 \in \monad(\makeset{m_1}), \hh, 
\goptwo_1 \in \monad(\makeset{m_1})$, 
to a generic effect $\gop \circ (\goptwo_1, \hh, \goptwo_n) \in 
\monad(\makeset{l})$, where 
$l \defeq \sum_{i \in \makeset{n}} m_i$. 
Additionally, operands comes with a diagrammatic syntax whereby we 
write 
$$
\eff{\gop}{n}{i}{x_i}
$$
for the pair $(\gop, \lan \varone_i \ran_{i \in \makeset{n}}) \in \monad(\makeset{n}) \times X^n$. 
In a diagram $\eff{\gop}{n}{i}{x_i}$, the letter 
$i$ ranges over elements in $\makeset{n}$ and to each $i$ it is associated the 
corresponding element $x_i$. That is, the horizontal bar with subscript $i$ and 
target $x_i$ stands for the function $i \mapsto x_i$.





\begin{example}
\begin{varenumerate}
\item Consider the maybe monad $\maybe$. We present an object
  $\mu \in \maybe(X)$ as a pair in 
  $\maybe(\makeset{n}) \times X^n$, 
  for some $n \in \mathbb{N}^{\infty}$. Since 
  $\maybe(\makeset{n}) \times X^n = 
  (\makeset{n} + \{\divergence\}) \times X^n$,
  $\mu$ is (presented as) either a pair 
  $(k, \lan x_i \ran_{i \in \makeset{n}})$ or a pair 
  $(\divergence, \lan x_i \ran_{i \in \makeset{n}})$. 
  The former corresponds to the case of convergence to $x_k$, whereas the latter 
  to divergence.
  In particular, if $\mtermone$ 
  is the result of evaluating a $\lambda$-term, then we will actually have 
  $n = 1$ (if the term converges) or $n = 0$ (if the term diverges).
  If $n = 1$, we obtain pairs of the form $(1, \lan x \ran)$), which we write as
  $\effconvergence{x}$.
  If $n=0$, then we can only have the pair $(\bot,\lan \ran)$, where 
  $\lan\ran$ is the empty sequence. 
  We write such a pair as $\boxed{\divergence}$.
\item Consider the output monad $\Output$. We present an object
  $\mu \in \Output(X)$ as a pair in 
  $\Output(\makeset{n}) \times X^n = 
  O^{\infty} \times (\makeset{n} + \{\divergence\}) \times X^n$, 
  for some $n \in \mathbb{N}^{\infty}$. Therefore, $\mu$ is presented as 
  either a triple $(o, \divergence, \lan x_i \ran_{i \in \makeset{n}})$, 
  or as a triple $(o, k, \lan x_i \ran_{i \in \makeset{n}})$. 
  The former case means that we have divergence, and that the string $o$ is outputted, 
  whereas the latter case means that we converge to $x_k$, and that the string $o$ 
  is outputted. 
  As before, if $\mu$ is the result of evaluating a ($\lambda$-)term, we will have 
  either $n = 1$ (if the term converges) or $n=0$ (if the term diverges). 
  If $n=1$, we have triples of the form $(o, 1,x)$) which we write as 
  $\effnull{o}{x}$. If $n=0$, the we can only have 
  triples of the form $(o,\bot,\lan\ran)$, which we write as 
  $\boxed{(o, \divergence)}$.
\item The case for the subdistribution monad goes exactly as in 
  Example~\ref{example:generic-effects-probabilistic-computations}. We can present a 
  formal sum $(\lan p_i \ran_{i \in \makeset{n}}, \lan x_i \ran_{i \in \makeset{n}})$ as 
  $\eff{\lan p_i \ran_{i \in \makeset{n}}}{n}{i}{x_i}$. 
\end{varenumerate}
\end{example}

Using the diagrammatic syntax, we present the composition of pairs 
$(\gop, \vect{x})$, $(\goptwo_1, \vect{y}_1), \hh, (\goptwo_n \vect{y}_n)$ 
(with $\vect{x}$ of length $n$ and each $\vect{y}_i$ of length $m_i$) as:
$$
\efftwo{\hspace{0.05cm}\gop_{\phantom i}}{n}{i}{\goptwo_i}{m_i}{j}{y_j}
$$
Notice that associativity of composition is built-in the diagrammatic notation
and that the latter also manages index dependencies. In fact, in the above 
diagram we see that
$i \in \makeset{n}$ and $j \in \makeset{m_i}$. Moreover, 
by reading from the right to the left we recover index dependencies:
since $j \in \makeset{m_i}$, it depends on $i \in \makeset{n}$. 
There is a trivial generic effect $\unit \in \monad(\makeset{1})$ 
corresponding to the unit of $\monad$
which behaves as a neutral element for composition: 
\begin{align*}
\shorteffunit{\mtermone} &= \mtermone;
&
\effshort{\gop}{n}{i}{\shorteffunit{x_i}} &= \effshort{\gop}{n}{i}{x_i}.
\end{align*} 
Moreover, given a function $f: X \to \monad(Y)$ and $\mu \in \monad(X)$, 
we see that if $\mu$ is presented as 
$\eff{\gop}{n}{i}{x_i}$, 
then $\mu \bind f$ is presented as 
$
\eff{\gop}{n}{i}{f(x_i)}.
$
Finally, we recall a well-known result by Plotkin and Power 
stating that algebraic operations and generic effects are equivalent 
notions.

\begin{proposition}[\cite{Plotkin/algebraic-operations-and-generic-effects/2003}]
\label{prop:alg-op-gen-eff}
There is a one-to-one correspondence between generic effects in $\monad(\makeset{n})$ and
$n$-ary algebraic operations on $\monad$.
\end{proposition}

In light of Proposition~\ref{prop:alg-op-gen-eff}, 
if we present objects $\mu_i \in \monad(X)$ as $\mtermone_i$, then 
we write $\eff{\op}{n}{i}{\mtermone_i}$ for the presentation 
of $\op(\mu_1, \hh, \mu_n)$.
Notice that if 
$\mu_i$ is presented as $\mtermone_i$, then both 
$\op(\mu_1, \hh, \mu_n) \bind f$ and $\op(\mu_1 \bind f, \hh, \mu_n \bind f)$ 
are presented as $\eff{\op}{n}{i}{f(\mtermone_i)}$. 
That is, the defining identity of algebraic operations 
(Definition~\ref{def:algebraic-operation}) is built-in the notation.

\paragraph{Order-Theoretic Properties}

Since we deal with $\signature$-continuous monads, we can transfer order-theoretic 
properties of $\monad$ to the diagrammatic notation by stipulating that 
a diagram $\mtermone$ is below a digram $\mtermtwo$ (notation 
$\mtermone \cpoleq \mtermtwo)$ if so are the elements presented by those diagrams.
In particular,
there is a bottom effect $\boteff \in \monad(\makeset{0})$ 
corresponding 
to the bottom element of $\monad$ satisfying the law 
$\boteff \cpoleq \mtermone$, for any diagram $\mtermone$. 
Additionally,we have the following monotonicity laws, where $f: X \to \monad(Z):$
\begin{align*}
(\forall i \in \makeset{n}.\ \mtermone_i \cpoleq \mtermtwo_i)
&\implies \effshort{\gop}{n}{i}{\mtermone_i} \cpoleq \effshort{\gop}{n}{i}{\mtermtwo_i}
\\
\effshort{\gop}{n}{i}{x_i} \cpoleq \effshort{\goptwo}{m}{j}{y_j}
&\implies 
\effshort{\gop}{n}{i}{f(x_i)} \cpoleq \effshort{\goptwo}{m}{j}{f(y_j)}
\end{align*}
}

\section{A Linear Calculus with Algebraic Effects}
\label{sect:environmental-equivalences}
In this section, we introduce 
a core \emph{linear} call-by-value calculus with 
\emph{algebraic operations} and \emph{explicit copying} and its 
\emph{resource-agnostic} operational semantics.
The syntax of the calculus is parametric with respect to 
a signature $\signature$ of operation symbols (notation $\op \in \signature$), 
whereas its dynamics 
relies on a $\signature$-continuous monad $\monad$, which we assume 
to be fixed.

\subsection{Syntax}
Our vehicle calculus is a linear refinement of 
fine-grain call-by-value \cite{Levy/InfComp/2003}, which we call $\Lambda^{!}$. 
The syntax of $\Lambda^{!}$ is given by two syntactic classes, 
\emph{values} (notation $\valone, \valtwo, \hh$) 
and \emph{computations} (notation $\termone, \termtwo, \hh$), 
which are thus defined: 
\begin{align*}
\valone &::= \varone 
\midd \abs{\varone}{\termone} 
\midd \bang{\termone}
\\
\termone &::= \evarone 
\midd \return{\valone} 
\midd \valone \valone 
\midd \seq{\termone}{\termone} 
\midd \op(\termone, \hh, \termone) 
\midd \coseq{\valone}{\termone}.
\end{align*}
\renewcommand{\eenvone}{\Sigma}
\renewcommand{\lenvone}{\Omega}
The letter $\varone$ denotes a \emph{linear} variable, and thus acts 
as a placeholder for a \emph{value} which has to be used exactly once. 
Dually, the letter $\evarone$ denotes a
\emph{non-linear} variable, and thus acts as a placeholder for a \emph{computation} 
which can be used \emph{ad libitum}. 

Following the fine-grain discipline, we require computations to be explicitly sequenced by means of the $\seq{-}{-}$ constructor. 
The latter comes in two flavors: in the first case, we deal with 
expressions of the form $\seq{\termone}{\termtwo}$, where $\varone$ is a \emph{linear}
variable in $\termtwo$ (and thus used once). The intuitive semantics of such an expression 
is to evaluate $\termone$, and then bind the result of the evaluation to $\varone$ 
in $\termtwo$. As $\varone$ is linear in $\termtwo$, the result of $\termone$ 
cannot be copied. 
In the second case, we 
deal with 
expressions of the form $\coseq{\valone}{\termtwo}$, where $\evarone$ is a \emph{non-linear}
variable in $\termtwo$ (and thus it can be used as will). As we are going to see, 
for such an expression to be meaningful, we need $\valone$ to be a banged 
computation $!\termone$.
The intuitive semantics of such an expression is thus
to `unbang' $!\termone$, and then bind $\termone$ to $\evarone$ 
in $\termtwo$, this way enabling $\termtwo$ to copy $\termone$ as will.

When the distinction between values and computations is not relevant, 
we generically refer to \emph{terms}, and denote them as $t, s, \hh$.
We adopt standard syntactic conventions as in \cite{Barendregt/Book/1984}.
In particular, we work with terms modulo renaming of bound variables,
and denote by $\subst{t}{\varone}{\valone}$ (resp. 
$\subst{t}{\evarone}{\termone}$) 
the result of capture-avoiding substitution 
of the value $\valone$ (resp. computation $\termone$) for the variable $\varone$ 
(resp. $\evarone$) in $t$.

\subsection{Statics} 
The syntax of $\Lambda^{\bang}$ allows one to write undesired programs, 
such as programs having runtime errors (e.g. $(\bang{\termone})\valone$) 
and programs that should be forbidden by any reasonable type system 
(such as $(\val{\bang{\termone}}) \oplus (\val{\abs{\varone}{\termtwo}})$). 
To overcome this problem, we follow \cite{CrubilleDalLago/ESOP/2017} and 
endow $\Lambda^{\bang}$ with a simply-typed system with 
recursive types, using the system in, e.g., 
\cite{DBLP:books/daglib/0032840}.
Types are 
defined by the following grammar:
\begin{align*}
\typeone &::= \mathsf{x} 
\midd ! \typeone
\midd \typeone \mmap \typeone 
\midd \mu \mathsf{x}. \typeone \mmap \typeone 
\midd \mu \mathsf{x}. !\typeone 
\end{align*}
where $\mathsf{x}$ is a type variable. Types are are defined up to equality, 
as defined in Figure~\ref{fig:type-equality}, where $\typeone[\typetwo/\mathsf{x}]$ denotes 
the substitution of $\typetwo$ for all the (free) occurrences of 
$\mathsf{x}$ in $\typeone$.

\begin{figure*}[htbp]
\hrule 
\[
\infer{\mu \mathsf{x}. \typeone \mmap \typetwo = 
\typeone[\mu \mathsf{x}. \typeone \mmap \typetwo/\mathsf{x}] \mmap
\typetwo[\mu \mathsf{x}. \typeone \mmap \typetwo/\mathsf{x}]}{}
\quad
\infer{\mu \mathsf{x}. !\typeone = 
 !\typeone[\mu \mathsf{x}. !\typeone/\mathsf{x}]}{}
\quad
\infer{\typeone = \typetwo}{
  \typeone = \rho[\typeone/\mathsf{x}]
  &
  \typetwo = \rho[\typetwo/\mathsf{x}]
}
\]
\hrule
\caption{Type equality}
\label{fig:type-equality}
\end{figure*} 

In order to define the collection of well-typed expressions, we consider 
sequents $\env{\eenvone}{\lenvone} \valimp \valone: \typeone$  
and $\env{\eenvone}{\lenvone} \compimp \termone: \typeone$, where 
$\lenvone$ is a linear environment, i.e. a set without repetitions of the form 
$\varone_1:\typeone_1, \hh, \varone_n:\typeone_n$, and $\eenvone$ is a 
\emph{non-linear} environment, i.e. a set without repetitions of the form 
$\evarone_1:\typetwo_1, \hh, \evarone_n: \typetwo_n$. 
Rules for derivable sequents are given in 
Figure~\ref{fig:lambda-bang-syntax}. We write 
$\values_{\typeone}$ and $\computations_{\typeone}$ for the 
collection of closed values and computations of type $\typeone$, respectively. 
We write $\values$ and $\computations$ when types are not relevant.

\begin{figure*}[htbp]
\hrule 
\[
\infer
  {\env{\eenvone}{\varone: \typeone} \valimp \varone: \typeone}{}
\quad
\infer
  {\env{\evarone: \typeone, \eenvone}{\emptyenv} \compimp \evarone:\typeone}{}
\quad
\infer
  {\env{\eenvone}{\lenvone} \valimp \abs{\varone}{\termone}: \typeone \mmap \typetwo}
  {\env{\eenvone}{\varone:\typeone, \lenvone} \compimp \termone: \typetwo}
\quad
\infer{\env{\eenvone}{\lenvone} \compimp \return{\valone}: \typeone}
{\env{\eenvone}{\lenvone} \valimp \valone: \typeone}
\]
\vspace{-0.1cm}
\[
\infer{\env{\eenvone}{\lenvone, \lenvone'} \compimp \valone \valtwo: \typetwo}
{\env{\eenvone}{\lenvone} \valimp \valone: \typeone \mmap \typetwo
& 
\env{\eenvone}{\lenvone'} \valimp \valtwo: \typeone}
\quad
\infer{\env{\eenvone}{\emptyenv} \valimp \bang{\termone}: \bang \typeone}
{\env{\eenvone}{\emptyenv} \compimp \termone: \typeone}
\quad 
\infer{\env{\eenvone}{\lenvone, \lenvone'} \compimp \coseq{\valone}{\termone}: \typetwo}
{\env{\eenvone}{\lenvone} \valimp \valone: \bang \typeone
& 
\env{\eenvone, \evarone: \typeone}{\lenvone'} \compimp \termone: \typetwo}
\]
\vspace{-0.1cm}
\[
\infer{\env{\eenvone}{\lenvone, \lenvone'} \compimp \seq{\termone}{\termtwo}: \typetwo}
{\env{\eenvone}{\lenvone} \compimp \termone: \typeone
&
\env{\eenvone}{\lenvone', \varone: \typeone} \compimp \termtwo: \typetwo}
\quad
\infer{\env{\eenvone}{\lenvone} \compimp \op(\termone_1, \hh, \termone_n): \typeone}
{\env{\eenvone}{\lenvone} \compimp \termone_1: \typeone 
& \hh & 
\env{\eenvone}{\lenvone} \compimp \termone_n: \typeone}
\]
\hrule
\caption{Statics of $\Lambda^!$}
\label{fig:lambda-bang-syntax}
\end{figure*}

\begin{remark}[Notational Convention]
In order to facilitate the communication 
of the main ideas behind this work and to lighten the (quite heavy) notation
we will employ in the next sections,
we avoid to mention types 
(and ignore them 
in the notation) whenever possible. 
Nonetheless, 
the reader should keep in mind
that from now on we work with typable terms only. 
We refer to such an assumption as the \emph{type assumption}. 
\end{remark}

\subsection{Dynamics} 
\label{sec:operational-semantics}
The dynamic semantics of $\Lambda^!$ associates to any 
\emph{closed computation} $\termone$ of type $\typeone$ a monadic 
element in $\monad(\values_{\typeone})$. 
\longv{
Such a dynamics is defined
relying on 
Felleisen's evaluation semantics 
\cite{Felleisen/SyntacticTheoriesSequentialControl/1992}. 
Accordingly, we define \emph{evaluation contexts} and 
\emph{redexes} by the following grammars
\begin{align*}
\ectxone &::= [-] 
\midd \seq{\ectxone}{\termone}
\\
\redex &::= (\abs{\varone}{\termone})\valone 
\midd \seq{(\return{\valone})}{\termone} 
\midd \coseq{\bang{\termone}}{\termtwo} 
\midd \op(\termone_1, \hh, \termone_n)
\end{align*}
where $[-]$ acts as a placeholder for a computation. 
The \emph{pure} reduction relation $\mapsto$ is thus defined:
\begin{align*}
(\abs{\varone}{\termone})\valone &\mapsto \subst{\termone}{\varone}{\valone} 
&
\seq{(\return{\valone})}{\termone} &\mapsto \subst{\termone}{\varone}{\valone}
&
\coseq{\bang{\termone}}{\termtwo} &\mapsto \subst{\termtwo}{\evarone}{\termone}
\end{align*}
Notice that $\mapsto$ is deterministic and that no (side) effect is produced when 
performing according $\mapsto$-reductions. We denote by $\redex'$
the unique term such that $\redex \mapsto \redex'$.
}
The dynamics of $\Lambda^!$ is defined
in Figure~\ref{fig:operational-semantics-lambda-bang} by means of an 
$\mathbb{N}$-indexed family of evaluation 
functions mapping a \emph{closed} 
computation $\termone \in \computations_{\typeone}$ 
to an element 
$\semln{\termone}{k} \in \monad(\values_{\typeone})$,
where we stipulate $\semln{\termone}{0} \defeq \bot$. 
Since $(\semln{\termone}{k})_{k \geq 0}$ forms an $\omega$-chain in 
$\monad (\values)$, we define 
$\seml{\termone} \defeq \lub_{k \geq 0} \semln{\termone}{k}$. 
Notice that thanks to the type assumption, we ignore programs causing runtime errors.
Finally, we lift $\seml{-}$ to monadic computations, i.e. 
to elements $\mstateone \in \monad (\Lambda)$ by 
setting 
$\semlstar{\mstateone} \defeq \mstateone \bind (\fun{\termone}{\seml{\termone}})$ 
(and similarity for $\semln{-}{k}$).

\begin{figure*}[htbp]

\hrule 
\begin{align*}
\semln{\val{\valone}}{k+1} 
&\defeq \unit(\valone)
\\
\semln{(\abs{\varone}{\termone})\valone}{k+1} &\defeq 
\semln{\subst{\termone}{\varone}{\valone}}{k} 
\\
\semln{\seq{\termone}{\termtwo}}{k+1} &\defeq 
  \semln{\termone}{k} \bind 
  (\fun{\valone}{\semln{\subst{\termtwo}{\varone}{\valone}}{k}})
\\
\semln{\coseq{\bang{\termone}}{\termtwo}}{k+1} 
  &\defeq \semln{\subst{\termtwo}{\evarone}{\termone}}{k}
\\
\semln{\op(\termone_1, \hh, \termone_n)}{k+1} 
&\defeq \sem{\op}(\semln{\termone_1}{k}, \hh, \semln{\termone_n}{k})
\end{align*}
\hrule
\caption{Operational Semantics of $\Lambda^{!}$}
\label{fig:operational-semantics-lambda-bang}
\end{figure*}

\subsection{Observational Equivalence} 
In order to compare $\Lambda^!$-terms, we introduce the notion
of \emph{contextual equivalence} 
\cite{Morris/PhDThesis}. 
To do so, we follow 
\cite{Gavazzo/ICTCS/2017,Simpson-Niels/Modalities/2018} and postulate that 
once an observer executes a program, 
she can only observe the effects produced by the evaluation of the program. 
For instance, in a pure (resp. probabilistic) calculus one 
observes pure (resp. the probability of) convergence. 
Following this postulate, we define an observation function
$\obslambdastar: \monad(\Lambda) \to \monad(1)$ as $\monad(!_{\values})$, where  
$1 = \{*\}$ is the one-element set and
$!_\values: \values \to 1$ is the terminal arrow. 
As a consequence, we see that $\obslambdastar$ is strict and continuous, 
so that we have, e.g., 
$\obslambdastar(\lub_k \mstateone_k) = \lub_k \obslambdastar(\mstateone_k)$.

\begin{example}
Notice that $\monad(1)$ indeed describes the observations 
one usually work with in concrete calculi. For instance, 
$\distribution(1) \cong [0,1]$, so that 
$\obslambdastar(\sem{\termone})$ gives the probability of convergence 
of $\termone$, and $\maybe(1) \cong \{\bot, \top\}$, so that 
$\obslambdastar(\sem{\termone}) = \top$ if and only if 
$\termone$ converges.
\end{example}

In order to define contextual equivalence, we need to introduce the notion 
of a $\Lambda^!$-context. The latter is simply a $\Lambda^{!}$-term 
with a single \emph{linear} hole $[-]$ acting as a placeholder for a computation 
(we regard a value 
$\valone$ as the computation $\val{\valone}$). 
We do not give an explicit definition of contexts, the latter being standard.

\shortv{
\begin{definition}
Define contextual equivalence $\ctxeq$ as follows:
\begin{align*}
\valone \ctxeq \valtwo &\iff \val{\valone} \ctxeq \val{\valtwo}
&
  \termone \ctxeq \termtwo &\iff \forall \ctxone.\ 
  \obslambdastar\ \sem{\ctxone[\termone]} = \obslambdastar\ \sem{\ctxone[\termtwo]}.
\end{align*}
\end{definition}
The universal quantification over contexts guarantees $\ctxeq$ 
to be a congruence relation. However, it also makes $\ctxeq$ difficult to
be used
in practice. 
We overcome this deficiency by characterising contextual 
equivalence as a suitable notion of trace equivalence.

}
\longv{
  \begin{definition}
  Define contextual equivalence $\ctxeq$ as follows:
  \begin{align*}
    \termone \ctxeq \termtwo &\iff \forall \ctxone.\ 
    \obslambdastar\ \sem{\ctxone[\termone]} = \obslambdastar\ \sem{\ctxone[\termtwo]}
  \\
   \valone \ctxeq \valtwo &\iff \val{\valone} \ctxeq \val{\valtwo}.
  \end{align*}
  \end{definition}
  As usual, we can easily show $\ctxeq$ to be a congruence relation. 

  \begin{remark}
  Thinking to a context $\ctxone$ as an experiment, we see that $\ctxone$ being 
  forced to use its hole $[-]$ \emph{linearly}, we are allowed to 
  experiment with a program $\termone$ more than once only if 
  $\termone \in \computations_{\bang \typeone}$. 
  \end{remark}

Contextual equivalence is a powerful notion to discriminate 
between programs, but are not well-suited to establish equivalences between 
them. We overcome this deficiency by characterising contextual 
equivalence as a notion of effectful environmental trace equivalence. 
}

\section{Resource-sensitive Semantics and Program Equivalence} 
\label{section:resource-sensitive-transition-systems}

The operational semantics of Section~\ref{sec:operational-semantics} 
is \emph{resource-agnostic}, meaning that linearity \emph{de facto} 
plays no role in the definition of the dynamics of a program. 
To overcome this deficiency, we endow $\Lambda^{!}$ with 
a resource-sensitive operational semantics: we give the latter 
by means of a suitable transition systems, which we dub
resource transition systems.
\emph{Resource transition systems} (RTSs, for short) 
provide an operational semantics for 
$\Lambda^!$-programs accounting for both their intensional and extensional behaviour. 
Those are defined as first-order transition 
systems in the spirit of \cite{DBLP:conf/concur/MadiotPS14}, and 
generalise the Markov chains of \cite{CrubilleDalLago/ESOP/2017}.

\subsection{Auxiliary Notions} 
In order to properly handle resources, 
it is useful to introduce some 
notation on sequences.  
Let $S, S'$ be sequences over objects $s_1, s_2, \hh$.
Unless ambiguous, we denote the 
concatenation of $S$ and $S'$ as 
$S,S'$. Moreover, for $S = s_1, \hh, s_k$ we denote by $|S| = k$
the length of $S$, and write $S[s]_i$, with $i \in \{1,\hh,k + 1\}$, 
for the sequence obtained by inserting $s$ in $S$ at position $i$, i.e. 
the sequence $s_1, \hh, s_{i-1}, s, s_i, \hh, s_k$ of length $k+1$.  
Given a sequence $S = s_1, \hh, s_k$, we will form new sequences out of it 
by taking elements in $S$ at given positions. If 
$\bar{c} = c_1, \hh, c_n$ is a sequence with elements in $\{1, \hh, k\}$ without 
repetitions,
    then we write $S_{\bar{c}}$ for the sequence $s_{c_1}, \hh, s_{c_n}$, and 
    $S \ominus \bar{c}$ for the sequence obtained from $S$ 
    by removing elements in positions $c_1, \hh, c_n$. 
    In order to preserve the order of $S$, we often consider sequences
    $\bar{c} = 
    (c_1 < \cc < c_n)$ with $c_i \in \{1, \hh, k\}$. We call such sequences valid for 
    $S$
    (although we should say valid for $|S|$
    ). 
\longv{
\begin{description}
  \item[Concatenation and insertion] 
    Unless ambiguous, we will denote the 
    concatenation of $\Sigma$ and $\Sigma'$ as 
    $\Sigma,\Sigma'$. Moreover, for $\Sigma = S_1, \hh, S_k$ we denote by $|\Sigma| = k$
    the length of $\Sigma$, and write $\Sigma[S]_p$, with $p \in \{1,\hh,k + 1\}$, 
    for the sequence obtained by inserting $S$ in $\Sigma$ at position $p$, i.e. 
    the sequence $S_1, \hh, S_{p-1}, S, S_p, \hh, S_k$ of length $k+1$. 
    Notice that $\Sigma[S]_1 = S, \Sigma$ and $\Sigma[S]_{k+1} = \Sigma, S$.  

  \item[Subsequences and subtraction] 
    Oftentimes, given a sequence $\Sigma = S_1, \hh, S_k$, we will form new sequences out of it 
    by taking elements in $\Sigma$ at given positions. If 
    $\bar{p} = (p_1, \hh, p_l)$ is a sequence with elements in $\{1, \hh, k\}$ without 
    repetitions,
    then we write $\Sigma_{\bar{p}}$ for the sequence $S_{p_1}, \hh, S_{p_l}$, and 
    $\Sigma \ominus \bar{p}$ for the sequence obtained from $\Sigma$ 
    by removing elements in position $p_1, \hh, p_l$. Observe that if 
    $\varphi: \{1, \hh, l\} \to \{1, \hh, l\}$ is a permutation, then 
    $\Sigma \ominus \bar{p} = \Sigma \ominus \varphi(\bar{p})$ (meaning that
    the operation $\ominus$ has actually a set---rather than a sequence---as left operand), 
    and that we have the identity $\Sigma[S]_p \ominus p = \Sigma$. 

  \item[Ordered subsequences]
    When building the sequence $\Sigma \ominus \bar{p}$, for $\Sigma$ and $p$ as above, 
    we preserve the order of 
    $\Sigma$.
    This is not the case for 
    $\Sigma_{\bar{p}}$.
    To avoid such behaviours, we can consider sequences $\bar{p} = 
    (p_1 < \cc < p_k)$ with $p_i \in \{1, \hh, n\}$. We call such sequences valid for 
    $\Sigma$
    (although we should say valid for $|\Sigma|$: indeed, 
    if $\bar{p}$ is valid for $\Sigma$, then it 
    is also valid for any $\Sigma'$ such that $|\Sigma'| = |\Sigma|$). 
\end{description}
}

\paragraph{System $\syskappa$}
The resource-sensitive operational semantics of $\Lambda^!$ is given by the RTS 
$\syskappa$.
Following
\cite{DBLP:conf/concur/MadiotPS14}, $\syskappa$-states are defined as 
\emph{configurations} $\econf{\gone}{\Theta}$, i.e.
pairs of sequences of terms, where 
$\gone$ is a 
(finite) sequence of (closed) computations and  
$\Theta$ is a (finite) sequence of (closed) terms 
in which only the last one need not be a value. 
\longv{In order to facilitate our analysis, we introduce the following notation. 
  If $\Theta$ ends with a closed computation $\termone$, then we 
  write $\econfterm{\gone}{\done}{\termone}$ with $\done$ finite 
  sequence of closed values (and $\Theta = \done, \termone$). Otherwise, 
  we write $\econf{\gone}{\done}$, with $\done$ as above.
}
To facilitate our analysis, we 
write $\econfterm{\gone}{\done}{\termone}$ if $\Theta = \done, \termone$, 
with $\done$ finite 
sequence of closed values and $\termone \in \computations$. 
Otherwise, we write $\econf{\gone}{\done}$, with $\done$ as above.

In a configuration $\econfterm{\gone}{\done}{\termone}$ (and similarity in 
$\econf{\gone}{\done}$), 
$\gone$
represents the non-linear resources available, which are 
(closed) computations: 
the environment can freely duplicate and evaluate them, as well as 
use them \emph{ad libitum} to build arguments to pass as input to 
other programs. Once a resource in $\gone$ has been used, it \emph{remains} 
in $\gone$, this way reflecting its non-linear nature.
Dually, $\done$
represents the linear resources available, which are closed values. 
Values in $\done$ being closed, they 
are either abstractions or banged computations. In the latter case, the environment can 
take a value $\bang{\termone}$, unbanged it, and put $\termone$ in $\gone$. 
In the former case, the environment can pass to a value $\abs{\varone}{\termtwo}$ 
an input argument made out of a context $\ctxone$ (provided by the very environment) 
using 
values and computations in $\gone, \done$. Since resources in $\done$ are linear,
once they are used by $\ctxone$, they must be removed from $\done$.
Finally, the program $\termone$ is the tested program. The environment can only 
evaluate it, possibly producing effects and values (linear resources). 
Once a linear resource $\valone$ has been produced, it is put in $\done$.

The calculus $\Lambda^!$ being typed, it is convenient to 
extend the notion of a type to configurations by defining a 
configuration type (notation $\alpha, \beta, \hh$) 
as a pair of sequences $\econf{\typeone_1, \hh, \typeone_n}{\typetwo_1, \hh, \typetwo_m}$ 
of ordinary types. 
We say that a
configuration $\econfone = \econf{\gone}{\Theta}$ has type 
$\alpha = \econf{\typeone_1, \hh, \typeone_n}{\typetwo_1, \hh, \typetwo_m}$ 
(and write $\imp \econfone: \alpha$) if 
each computation $\termone_i$ at position $i$ in $\gone$ has type 
$\typeone_i$, and each term $t_i$ at position $i$ in $\Theta$ has 
type $\typetwo_i$. 

Notice that configuration types almost completely describe the structure of 
configurations. However, they do not allow one to see whether 
the last argument in the second component $\Theta$ of a configuration 
$\econf{\gone}{\Theta}$ is a value 
(so that the type will be inhabitated by configurations of the form 
$\econf{\gone}{\done}$) or a computation 
(so that the type will be inhabitated by configurations of the form 
$\econfterm{\gone}{\done}{\termone}$). To avoid this issue, we add a special label 
to the last type $\typetwo_m$ of the second component of a configuration type, 
this way specifying whether $\typetwo_m$ refers to a value or to a computation. 
\longv{
  Mimicking 
  previous notational conventions, we write 
  $\econf{\typeone_1, \hh, \typeone_n}{\typetwo_1, \hh, \typetwo_m}$ if all 
  $\typetwo_i$s refer to values, and
  $\econfterm{\typeone_1, \hh, \typeone_n}{\typetwo_1, \hh, \typetwo_m}{\typethree}$
  if all $\typetwo_i$s refers to values and $\typethree$ to a computation.

  Formally:
  \[
  \infer{\imp \econf{\termone_1, \hh, \termone_n}{\valone_1, \hh, \valone_m}:
  \econf{\typeone_1, \hh, \typeone_n}{\typetwo_1, \hh, \typetwo_m}}
  {\compimp \termone_1, \hh, \termone_n: \typeone_1, \hh, \typeone_n 
  & 
  \valimp \valone_1, \hh, \valone_m: \typetwo_1, \hh, \typetwo_m}
  \]
  \[
  \infer{\imp \econfterm{\termone_1, \hh, \termone_n}{\valone_1, \hh, \valone_m}{\termone}:
  \econfterm{\typeone_1, \hh, \typeone_n}{\typetwo_1, \hh, \typetwo_m}{\typethree}}
  {\compimp \termone_1, \hh, \termone_n: \typeone_1, \hh, \typeone_n 
  & 
  \valimp \valone_1, \hh, \valone_m: \typetwo_1, \hh, \typetwo_m
  & 
  \compimp \termone: \typethree}
  \]
}

We denote by $\envconf_{\typeconfone}$ the collection of configurations 
of type $\typeconfone$.
Notice that if $\econfone, \econftwo \in \envconf_{\typeconfone}$, then 
they have the same structure. In particular, terms in 
$\econfone$ and $\econftwo$ at the same position have the same type 
and belong to the same syntactic class.
As usual, following the type assumption, we will omit configuration 
types whenever possible.

States of $\syskappa$ are thus (typable) configurations, whereas its dynamics 
is based on three kind of actions: 
\emph{evaluation}, \emph{duplication}, and 
\emph{resource-based application}, 
which are \emph{extensional}, \emph{intensional}, and 
\emph{mixed extensional-intensional} actions, respectively. 
Formally,
we consider transitions from (typable) configurations, i.e. elements in 
$\bigcup_{\typeconfone} \envconf_{\typeconfone}$ to monadic configurations 
in $\bigcup_{\typeconfone} \monad(\envconf_{\typeconfone})$, i.e. 
monadic configurations $\mconfone$ such that all configurations 
in the support of $\mconfone$ have the same type. This ensures that 
all configurations in $\support{\mconfone}$ can make the same actions. 
As usual, such a property follows by typing, hence by the type assumption. 
We now spell out the main ideas behind the dynamics of 
$\syskappa$.
\begin{varitemize}
\item
Given a configuration 
$\econfterm{\gone}{\done}{\termone}$, the environment simply evaluates $\termone$. 
That is, we have the transition:
$$
\econfterm{\gone}{\done}{\termone} \xrightarrow{\eval} 
\sem{\termone} \bind (\fun{\valone}{\unit\econf{\gone}{\done, \valone}}).
$$
\item
Given a configuration of the form $\econf{\gone}{\done[\bang{\termone}]_l}$, 
the environment adds $\termone$ to the non-linear environment, 
and removes $\bang{\termone}$ 
from the linear one. We thus have the transition:
    $$
    \econf{\gone}{\done[\bang{\termone}]_l} \xrightarrow{\ ?_l\ }
    \unit \econf{\gone, \termone}{\done}.
    $$
\item
In a configuration of the form 
    $\econf{\gone[\termone]_l}{\done}$, the environment 
    has the non-linear resource $\termone$
    at its disposal, which can be duplicated (and eventually evaluated via an $\eval$ action).
    We model such a behaviour as the following transition (notice that $\termone$ is not removed from $\gone[\termone]_l$):
    $$
    \econf{\gone[\termone]_l}{\done} \xrightarrow{\ !_l\ } 
    \unit \econfterm{\gone[\termone]_l}{\done}{\termone}.
    $$
\item
 For the last action, namely resource-based application, we consider 
    open terms as playing the role of contexts. 
    An open term is simply a term 
    $\env{\eenvone}{\lenvone} \imp t$.
    We refer to an open term 
    $\env{\evarone_1, \hh, \evarone_n}{\varone_1, \hh, \varone_m} 
    \imp t$ as a $(n,m)$-(value/computation) context, 
    depending on whether $t$ is a value or a computation. 
    Given 
    sequences $\gone = \termone_1, \hh, \termone_n$, $\done = \valone_1, \hh, \valone_m$, 
    we write $t[\gone,\done]$ for the substitution of variables in $t$
    with the corresponding elements in 
    $\gone, \done$.
    As usual, following the type-assumption we assume types of variables to match types 
    of the substituted terms. 
    Given sequences $\bar{\imath}$, $\bar{\jmath}$ of length $n$, $m$ 
    valid for $\gone$, $\done$, respectively, we can build a new 
    (closed) term out of $\gone, \done$ 
    and a $(n,m)$-context $t$ as $t[\gone_{\bar{\imath}}, \done_{\bar{\jmath}}]$. 
    Since resources in $\done$ are linear, the construction of 
    $t[\gone_{\bar{\imath}}, \done_{\bar{\jmath}}]$ affects $\done$, this 
    way leaving only resources $\done \ominus \bar{\jmath}$ available. 
    We formalise this behaviour as the transition:
    \[
    \infer{\econf{\gone}{\done[\abs{\varone}{\termtwo}]_l} 
    \xrightarrow{ (\bar{\imath},\bar{\jmath},l ,t)} 
    \unit \econfterm{\gone}{\done \ominus \bar{\jmath}}
    {\subst{\termtwo}{\varone}{t[\gone_{\bar{\imath}}, \done_{\bar{\jmath}}]}}}
    {t\ (n,m)\text{-value context}
    & 
    |\bar{\imath}| = n, |\bar{\jmath}| = m
    &
    \bar{\imath}, \bar{\jmath} \text{ valid for } \gone, \done
    }
    \]
\end{varitemize}

\begin{definition}
\label{def:system-k}
System $\syskappa$ 
is the (resource) transition system having typable configurations as 
states, actions 
$$
\{\eval, ?_l, !_l, , (\bar{\imath},\bar{\jmath}, l, t), \typeconfone \mid 
l \in \mathbb{N}, t\ (n,m)\text{-value context}, |\bar{\imath}| = n, 
|\bar{\jmath}|=m \}
$$
where $\typeconfone$ ranges over configuration types,
and dynamics defined by the transition rules  
in Figure~\ref{fig:transition-environmental-lts}, where 
we employ the notation of previous discussion.
\end{definition}

\begin{figure*}[htbp]
\hrule 
\vspace{0.2cm}
\begin{align*}
\econfterm{\gone}{\done}{\termone} 
&\xrightarrow{\eval} 
\sem{\termone} \bind \fun{\valone}{\unit\econf{\gone}{\done, \valone}}
& 
\econf{\gone}{\done[\bang{\termone}]_l} 
&\xrightarrow{\ ?_l\ }
\unit \econf{\gone, \termone}{\done}.
\\
\econf{\gone[\termone]_l}{\done} 
&\xrightarrow{\ !_l\ } 
    \unit \econfterm{\gone[\termone]_l}{\done}{\termone}
& 
\econf{\gone}{\done[\abs{\varone}{\termtwo}]_l} 
&\xrightarrow{ (\bar{\imath},\bar{\jmath},l ,t)} 
    \unit \econfterm{\gone}{\done \ominus \bar{\jmath}}
    {\subst{\termtwo}{\varone}{t[\gone_{\bar{\imath}}, \done_{\bar{\jmath}}]}}
\end{align*}
\vspace{-0.2cm}
\hrule
\caption{Transition rules for $\syskappa$}
\label{fig:transition-environmental-lts}
\end{figure*}


\begin{remark}
\label{rem:well-typed-configurations}
Notice that given $\econfone \in \envconf_{\typeconfone}$, 
$\econfone$ can always make a $\typeconfone$-transition, this way making 
its type visible. Additionally, we see that
the transition structure of $\syskappa$ is \emph{type-driven}. That is,
given a configuration $\econfone \in \envconf_{\typeconfone}$ 
and a $\syskappa$-action $\ell$, 
$\typeconfone$ and $\ell$ alone determine whether $\econfone$ can make 
an $\ell$-transition. Moreover, if that is the case, then there is a unique 
$\mconfone$ such that 
$\econfone \xrightarrow{\ \ell\ } \mconfone$. Besides,
$\mconfone \in \monad(\envconf_{\typeconftwo})$ for some configuration type 
$\typeconftwo$ which is \emph{uniquely} determined by $\ell$ and $\typeconfone$. 
That is, there is a \emph{partial} function $\typefun$ from configuration 
types and
actions such that if $\typefun(\typeconfone, \ell)$ is defined and 
$\econfone \in \envconf_{\typeconfone}$, then $\econfone \xrightarrow{\ \ell\ } \mconfone$ 
with $\mconfone \in \monad(\envconf_{\typefun(\typeconfone,\ell)})$.
\longv{As a consequence, in order to know whether a configuration $\econfone$ of type 
$\typeconfone$ can make a $\ell$-transition, it is sufficient to check if 
$\typefun(\typeconfone,\ell)$ is defined.} 
From now on, we write $\typefun(\typeconfone,\ell) = \typeconftwo$ to mean 
that $\typefun(\typeconfone,\ell)$ is defined and equal $\typeconftwo$. As a consequence, 
we have the rule:
\[
\econfone \in \envconf_{\typeconfone} \wedge \typefun(\typeconfone,\ell) = \beta
\implies \exists! \mconfone \in \monad(\envconf_{\typeconftwo}).\ 
  \econfone \xrightarrow{\ \ell\ } \mconfone.
\]
\end{remark}

Having defined system $\syskappa$, there are at least two natural ways to 
compare its states. The first one is by means of \emph{bisimilarity}, which can 
be defined in a standard way \cite{DalLagoGavazzoLevy/LICS/2017}. 
Unfortunately, bisimilarity being sensitive to branching, 
it is bound not to work well for our purposes, 
as already extensively discussed.
The second natural way to compare $\syskappa$-states is by means of
\emph{trace equivalence} which, contrary to bisimilarity, 
is not sensitive to branching, and thus 
qualifies as a suitable candidate program equivalence for our purposes. 

\begin{definition}
A $\syskappa$-trace (just trace) is a finite sequence of $\syskappa$-actions. 
That is, a trace $\traceone$ is either the empty sequence (denoted by $\varepsilon$), 
or a sequence of the form $\ell \cdot \tracetwo$, where $\ell$ is a $\syskappa$-action 
and $\tracetwo$ a trace.
\end{definition}

We are interested in observing the behaviour of $\syskappa$-states 
on those traces that are coherent with their type. Therefore, given 
a $\syskappa$-state $\econfone$, we define the set $Tr(\econfone)$  of its 
traces by stipulating that $\varepsilon \in Tr(\econfone)$, for any $\econfone$, 
and that $\ell \cdot \tracetwo \in Tr(\econfone)$ whenever
$\econfone \xrightarrow{\ \ell\ } \mconfone$, for some monadic configuration 
$\mconfone$, and $\tracetwo \in Tr(\econftwo)$, for any $\econftwo \in 
\support{\mconfone}$. Notice that the latter clause is meaningful, since 
$Tr(\econfone)$ is actually determined by the type of $\econfone$ (rather than 
by $\econfone$ itself), and if $\econfone \xrightarrow{\ \ell\ } \mconfone$, 
then all configurations in the support of $\mconfone$ have the same type.

Now, given a $\syskappa$-state $\econfone$, and a trace $\traceone \in Tr(\econfone)$, 
the observable behaviour of $\econfone$ on $\traceone$ is
the element in $\monad(1)$ computed using the map $\sat$ thus defined:
\begin{align*}
\sat(\econfone, \varepsilon) &\defeq \unit(*);
&
\sat(\econfone, \ell \cdot \tracetwo) &\defeq 
  \mconfone \bind (\fun{\econftwo}{\sat(\econftwo, \tracetwo)}) 
   \text{ where } \econfone \xrightarrow{\ \ell\ } \mconfone.
\end{align*}

\begin{example}
\longv{
It is a straightforward exercise to prove that on the powerset monad 
$\sat$ gives the usual notion of `passing a trace'. }
Let us consider the 
(sub)distribution monad $\mathcal{D}$, and let $\econfone$ be a configuration. 
Recall that $\mathcal{D}(1) \cong [0,1]$, and notice that 
$\sat(\econfone, \varepsilon) = 1$. Suppose now 
$\econfone \xrightarrow{\eval} \sum_{i \in n} p_i \cdot \econftwo_i$. 
Then, we see that 
$\sat(\econfone, \eval \cdot \tracetwo) = 
\sum_{i \in n} p_i \cdot \sat(\econftwo_i, \tracetwo) \in [0,1]$, meaning 
that $\sat(\econfone, \traceone)$ gives the probability that 
$\econfone$ passes the trace $\traceone$.
\end{example}

\begin{definition}
The relation $\kappatraceeq$ on $\syskappa$-states is thus defined:
$$
\econfone \kappatraceeq \econftwo \iff Tr(\econfone) = Tr(\econftwo) 
\wedge
\forall \traceone \in Tr(\econfone).\ \sat(\econfone,\traceone) = 
\sat(\econftwo, \traceone)
$$
\end{definition}
 
We extend
the action of $\kappatraceeq$ to $\Lambda^!$-terms 
by regarding a computation $\termone$ as the 
configuration $\econfterm{\emptyenv}{\emptyenv}{\termone}$, 
and a value $\valone$ as the computation $\val{\valone}$. We denote the resulting notion 
$\lambdatraceeq$.

Having added $\kappatraceeq$ to our arsenal 
of operational techniques, it is time to investigate its structural properties 
and its relationship with contextual equivalence. Before doing so, 
however, we take a fresh look at our running example. 

\begin{example}
Let us use the machinery developed so far to review our introductory examples. 
First, we show 
$$
\return{\abs{\varone}{(\termone \oplus \termtwo)}}
\lambdatraceeq (\return{\abs{\varone}{\termone}}) \oplus 
(\return{\abs{\varone}{\termtwo}}).
$$
Let us call $\termthree$ the former program, and $\termfour$ the latter.
To see that $\termthree \lambdatraceeq \termfour$, we simply observe that 
$Tr\econfterm{\emptyenv}{\emptyenv}{\termthree} = 
Tr\econfterm{\emptyenv}{\emptyenv}{\termfour}$ 
and that for any $\traceone \in Tr(\termthree)$, the probability that 
$\econfterm{\emptyenv}{\emptyenv}{\termthree}$ 
passes $\traceone$ coincides with the one of 
$\econfterm{\emptyenv}{\emptyenv}{\termfour}$. 
All of this can be easily observed by inspecting the following transition 
systems.
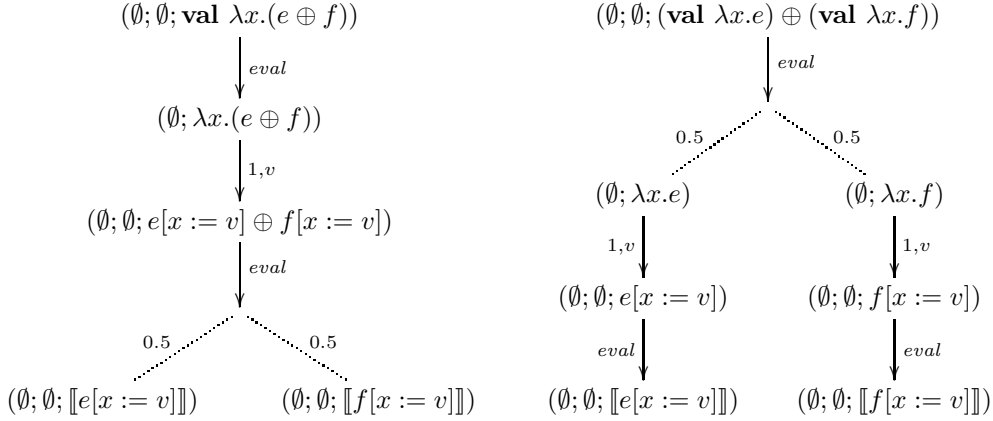
\[
\xymatrix@R=0.8cm@C=-4pc{
          & \econfterm{\emptyenv}{\emptyenv}{
          \return{\abs{\varone}{(\termone \oplus \termtwo)}}}
          \ar[d]^{\eval} & 
          \\
          & \econf{\emptyenv}{
          \abs{\varone}{(\termone \oplus \termtwo)}} 
          \ar[d]^{1,\valone} &
          \\
          & \econfterm{\emptyenv}{\emptyenv}{
          \subst{\termone}{\varone}{\valone} 
          \oplus 
          \subst{\termtwo}{\varone}{\valone}} 
          \ar[d]^{\eval}
          &
          \\
          & \ar@{.}[ld]_{0.5} 
            \ar@{.}[rd]^{0.5} 
          & 
          \\
          \econfterm{\emptyenv}{\emptyenv}
          {\sem{\subst{\termone}{\varone}{\valone}}}      
          &   & 
          \econfterm{\emptyenv}{\emptyenv}
          {\sem{\subst{\termtwo}{\varone}{\valone} }}    
          }
\qquad 
\xymatrix@R=0.8cm@C=-5pc{
          & 
          \econfterm{\emptyenv}{\emptyenv}{
          (\return{\abs{\varone}{\termone}}) \oplus 
          (\return{\abs{\varone}{\termtwo}})}
          \ar[d]^{\eval} 
          &
          \\
          & \ar@{.}[ld]_{0.5} \ar@{.}[rd]^{0.5} & 
          \\
          \econf{\emptyenv}{
          \abs{\varone}{\termone}}
          \ar[d]_{1,\valone}
          & &
          \econf{\emptyenv}{\abs{\varone}{\termtwo}}
          \ar[d]^{1,\valone} 
          \\
          \econfterm{\emptyenv}{\emptyenv}{
          \subst{\termone}{\varone}{\valone}}
          \ar[d]_{\eval}
          & &
          \econfterm{\emptyenv}{\emptyenv}{
          \subst{\termtwo}{\varone}{\valone}}
          \ar[d]^{\eval}
          \\
          \econfterm{\emptyenv}{\emptyenv}
          {\sem{\subst{\termone}{\varone}{\valone}}}      
          &   & 
          \econfterm{\emptyenv}{\emptyenv}
          {\sem{\subst{\termtwo}{\varone}{\valone} }}  
}
\]
In light of Theorem~\ref{thm:full-abstraction}, we can then conclude 
$\termthree \ctxeq \termfour$.
Next, we prove that such an equivalence is only linear:
$
\return{\bang{(\termone \oplus \termtwo)}}
\not \ctxeq
(\return{\bang{\termone}}) \oplus (\return{\bang{\termtwo}})
$.
For that, it is sufficient to instantiate $\termone$ and $\termtwo$ 
as the identity program $\return{(\abs{\varone}{\return{\varone}})}$
and the purely divergent program $\Omega$, respectively, 
and to take the context $\ctxone$ defined as
$\seq{[-]}{\coseq{\varone}{(\evarone;\evarone;\return{\valone})}}$,
where $\valone$ is closed value, and $\termone;\termtwo$ denotes 
trivial sequencing. Indeed, what $\ctxone$ does is to evaluate its 
input and then test the result thus obtained \emph{twice}.
\end{example}

\shortv{
\subsection{Full Abstraction of Trace Equivalence} 
\label{subsection:soundness_and_completeness}

In this section, we outline the proof of
\emph{full abstraction} of trace equivalence  
for contextual equivalence. 
Our proof of full abstraction builds upon the technique 
given by Deng and Zhang \cite{DBLP:journals/tcs/DengZ15} and 
Crubill\'e and Dal Lago \cite{CrubilleDalLago/ESOP/2017} 
to prove similar full abstraction results for trace equivalences and metrics, 
respectively. 
Due to the large 
amount of technicalities, the full proof of full abstraction of trace equivalence  
goes beyond the scope of this paper, so that here we only outline its 
main points.
Let us begin by showing that trace equivalence is \emph{sound} 
for contextual equivalence. 
\begin{proposition}
\label{thm:soundness}
${\lambdatraceeq} \subseteq {\ctxeq}$.
\end{proposition}
To prove Proposition~\ref{thm:soundness}, we have to show that 
if 
$\termone \lambdatraceeq \termtwo$, then we have
$\obslambdastar \seml{\ctxone[\termone]} = \obslambdastar \seml{\ctxone[\termone]}$, 
for any context $\ctxone$. 
Our proof proceeds by progressively building systems 
with increasingly more complex state spaces, but with finer dynamics. 
We summarise our strategy in the following diagram. 
$$
\xymatrix@C=2.5pc@R=1.5pc{
  \syslambda \ar[rrr]^{\ctxone[-]} \ar@{^{(}->}[d] 
  & 
  &
  & \strongkleisli{\syslambda} \ar[r]^{\obslambdastar} 
  & \monad 1
  \\
  \syskappa \ar@{^{(}->}[r]
  & \strongkleisli{\syskappa} \ar[r]^{\ctxone[-]}
  & \sysf \ar@{^{(}->}[r]
  & \strongkleisli{\sysf} \ar[u]^{\push} \ar[ru]_{\obsfstar}
  &
}
$$
Since $\lambdatraceeq$ is defined in terms of $\kappatraceeq$, we consider 
configurations---$\syskappa$-states---and contexts for them, where a context 
for a $\syskappa$-state $\econfone$ is 
just a standard multiple-holes context whose holes have to be filled with with 
terms in $\econfone$.
The first step of our strategy 
is the \emph{determinization} of $\syskappa$. This is achieved 
by lifting the state space of $\syskappa$ from configurations to monadic 
configurations. The dynamics of $\syskappa$ is then lifted relying on the (strong) 
monad structure of $\monad$ in a standard way \cite{DBLP:journals/tcs/LagoGT20}. 
We call the resulting system $\strongkleisli{\syskappa}$. 
The advantage of working with $\strongkleisli{\syskappa}$ is that 
$\strongkleisli{\syskappa}$-bisimilarity and 
$\strongkleisli{\syskappa}$-trace equivalence coincide, 
$\strongkleisli{\syskappa}$ being 
deterministic. 
In general, most of the transition systems we rely on can be ultimately described 
as systems $\mathcal{S} = (X, \delta)$ made of a state space $X$ and 
a dynamics $\delta: X \to \monad(X)^A$, for some set $A$ of actions. 
The determinization of $\mathcal{S}$, which we usually denote by 
$\mathcal{S}^*$, has $\monad(X)$ as state space and dynamics 
$\delta^*: \monad(X) \to \monad(X)^A$ defined as the strong 
Kleisli extension of $\delta$ (modulo (un)currying).

Having determinized $\syskappa$, we reach a situation where we have to study the computational behaviour 
of a monadic configuration $\mconfone$ --- 
i.e. a $\strongkleisli{\syskappa}$-state --- and 
a context $\ctxone$ for the configurations in the support of $\mconfone$.  
To do so, we build a further system, called $\sysf$, whose states are pairs 
$\pairconf{\ctxone}{\mconfone}$ made of a monadic configuration $\mconfone$ 
and a context $\ctxone$ for it. 
The dynamics of $\sysf$ is given by an evaluation function which, when applied to a 
$\sysf$-state 
$\pairconf{\ctxone}{\mconfone}$, gives the 
same result of evaluating the \emph{monadic computation} 
$\ctxone[\mconfone] \in \monad (\computations)$, 
where $\ctxone[\mconfone] = \mconfone \bind 
(\fun{\econfone}{\unit(\ctxone[\econfone])})$.
Such a dynamics explicitly separates the computational 
steps acting on $\ctxone$ only from those making $\ctxone$ and $\mconfone$ interact. 
This feature is crucial, as it shows that any interaction between $\ctxone$ and 
$\mconfone$ corresponds 
to a $\strongkleisli{\syskappa}$-action, so that equivalent 
$\strongkleisli{\syskappa}$-states will have the same $\sysf$-dynamics when paired 
with the same context. 
That gives us a finer analysis of the computational behaviour of 
the compound monadic computation $\ctxone[\mconfone]$, and ultimately of a 
compound computation $\ctxone[\termone]$. 
As we did for $\syskappa$, it is actually convenient to determinise $\sysf$. 
We call the resulting system $\strongkleisli{\sysf}$.
Finally, from $\strongkleisli{\sysf}$ we can come back 
to $\monad(\Lambda)$ 
using the map $\push: \strongkleisli{\sysf} \to \monad(\Lambda)$ 
defined by $\push(\xi) \defeq \xi 
\bind (\pairconf{\ctxone}{\kappa} \mapsto \ctxone[\kappa])$.
We summarize the systems introduce in the following table.

\begin{center}
\begin{tabular}{|ccccc|}
\hline
\textbf{System} & $\syskappa$ & $\strongkleisli{\syskappa}$ & $\sysf$ & $\strongkleisli{\sysf}$
\\
\textbf{States} 
& Configurations $K$ 
& Monadic configurations $\kappa$
& Pairs $\pairconf{\ctxone}{\mconfone}$ 
& Monadic pairs
\\
\textbf{Dynamics} 
& Definition~\ref{def:system-k}
& Kleisli lifting of $\syskappa$
& $\strongkleisli{\sem{\ctxone[\mconfone]}}$ 
& Kleisli lifting of $\sysf$
\\
\hline
\end{tabular}
\end{center}

What remains to be clarified is how relations between computations 
can be transformed into relations on the aforementioned systems. The answer to 
this question is given by the following \emph{lax}\footnote{
  Each square gives a set-theoretic inclusion. For instance, the leftmost 
  square states that ${\lambdatraceeq} \subseteq {\kappatraceeq}$.
} commutative diagram:
$$
\xymatrixcolsep{3pc}
\xymatrix{
  \syslambda \ar@{^{(}->}[r] \ar[d]_{\lambdatraceeq}|-*=0@{|}
  & \syskappa \ar@{^{(}->}[r] \ar[d]_{\kappatraceeq}|-*=0@{|}
  & \strongkleisli{\syskappa} \ar[r]^{\ctxone[-]} 
    \ar[d]_{\kappastartraceeq}|-*=0@{|}
  & \sysf \ar@{^{(}->}[r]
    \ar[d]_{\mathtt{C}(\kappastartraceeq)}|-*=0@{|}
  & \strongkleisli{\sysf} \ar[r]^{\obsfstar} 
    \ar[d]_{\bc{\kappastartraceeq}}|-*=0@{|}
  & \monad 1 \ar[d]_{=}|-*=0@{|}
  \\
  \syslambda \ar@{^{(}->}[r]
   & \syskappa \ar@{^{(}->}[r]
  & \strongkleisli{\syskappa} \ar[r]_{\ctxone[-]} 
  & \sysf \ar@{^{(}->}[r]
  & \strongkleisli{\sysf} \ar[r]_{\obsfstar} 
  & \monad 1
} 
$$
Here, $\mathtt{C}(\relone)$ denotes the contextual closure of $\relone$, whereas 
$\mathtt{B}(\relone)$ is the Barr extension 
of $\relone$ \cite{Barr/LMM/1970,Kurz/Tutorial-relation-lifting/2016}. 
Finally, the map $\obsfstar$ is obtained postcomposing the observation map $\obs$ 
with $\push$.
Let us now move to full abstraction.
\begin{theorem}
\label{thm:full-abstraction}
  ${\ctxeq} = {\lambdatraceeq}$.
\end{theorem}

To prove Theorem~\ref{thm:full-abstraction} it is 
sufficient to show ${\ctxeq} \subseteq {\lambdatraceeq}$. 
The latter is proved by noticing that any $\syskappa$-action can be encoded as 
a context. The encoding of $\syskappa$-actions 
as contexts is essentially the same one
of the one given by Crubill\'e and Dal Lago \cite{CrubilleDalLago/ESOP/2017} 
and thus we refer to the latter for details.
}

\longv{
\section{Trace Equivalence: Soundness and Completeness}
\label{sect:soundness-and-completeness}

In this section, we prove the main result of this work, namely 
\emph{full abstraction} of trace equivalence  
for contextual equivalence: ${\lambdatraceeq} =  {\ctxeq}$. 
That $\ctxeq$ is included in $\lambdatraceeq$ (\emph{completeness}) 
does not come with much of a surprise. In fact, 
it is easy to realise that all $\syskappa$-actions 
(and thus traces) can be implemented by suitable contexts 
\cite{CrubilleDalLago/ESOP/2017}. 
Proving that $\lambdatraceeq$ is included in $\ctxeq$ (i.e. \emph{soundness}) is, however, more 
challenging.
Our proof builds upon the technique 
given by Deng and Zhang \cite{DBLP:journals/tcs/DengZ15} and 
Crubill\'e and Dal Lago \cite{CrubilleDalLago/ESOP/2017} 
to prove similar full abstraction results for trace equivalences and metrics, 
respectively. 
Due to the large 
amount of technicalities, before entering into the technical details of the proof 
of soundness of trace equivalence, 
it is instructive to outline the 
main points of such a proof.

Soundness of trace equivalence means that the inclusion
${\lambdatraceeq} \subseteq {\ctxeq}$ holds. To prove that, 
we have to show that 
if 
$\termone \lambdatraceeq \termtwo$, then we have
$\obslambdastar \seml{\ctxone[\termone]} = \obslambdastar \seml{\ctxone[\termone]}$, 
for any context $\ctxone$. 
Our proof proceeds by progressively building systems 
with increasingly more complex state spaces, but with finer dynamics. 
We summarise our strategy in the following diagram. 
$$
\xymatrix@C=2.5pc@R=1.5pc{
  \syslambda \ar[rrr]^{\ctxone[-]} \ar@{^{(}->}[d] 
  & 
  &
  & \strongkleisli{\syslambda} \ar[r]^{\obslambdastar} 
  & \monad 1
  \\
  \syskappa \ar@{^{(}->}[r]
  & \strongkleisli{\syskappa} \ar[r]^{\ctxone[-]}
  & \sysf \ar@{^{(}->}[r]
  & \strongkleisli{\sysf} \ar[u]^{\push} \ar[ru]_{\obsfstar}
  &
}
$$
Since $\lambdatraceeq$ is defined in terms of $\kappatraceeq$, we consider 
configurations---$\syskappa$-states---and contexts for them, where a context 
for a $\syskappa$-state $\econfone$ is 
just a standard multiple-holes context whose holes have to be filled with with 
terms in $\econfone$.
The first step of our strategy 
is the \emph{determinization} of $\syskappa$. This is achieved 
by lifting the state space of $\syskappa$ from configurations to monadic 
configurations. The dynamics of $\syskappa$ is then lifted relying on the (strong) 
monad structure of $\monad$ in a standard way \cite{DBLP:journals/tcs/LagoGT20}. 
We call the resulting system $\strongkleisli{\syskappa}$. 
The advantage of working with $\strongkleisli{\syskappa}$ is that 
$\strongkleisli{\syskappa}$-bisimilarity and 
$\strongkleisli{\syskappa}$-trace equivalence coincide, 
$\strongkleisli{\syskappa}$ being 
deterministic. 
In general, most of the transition systems we rely on can be ultimately described 
as systems $\mathcal{S} = (X, \delta)$ made of a state space $X$ and 
a dynamics $\delta: X \to \monad(X)^A$, for some set $A$ of actions. 
The determinization of $\mathcal{S}$, which we usually denote by 
$\mathcal{S}^*$, has $\monad(X)$ as state space and dynamics 
$\delta^*: \monad(X) \to \monad(X)^A$ defined as the strong 
Kleisli extension of $\delta$ (modulo (un)currying).

Having determinized $\syskappa$, we reach a situation where we have to study the computational behaviour 
of a monadic configuration $\mconfone$ --- 
i.e. a $\strongkleisli{\syskappa}$-state --- and 
a context $\ctxone$ for the configurations in the support of $\mconfone$.  
To do so, we build a further system, called $\sysf$, whose states are pairs 
$\pairconf{\ctxone}{\mconfone}$ made of a monadic configuration $\mconfone$ 
and a context $\ctxone$ for it. 
The dynamics of $\sysf$ is given by an evaluation function which, when applied to a 
$\sysf$-state 
$\pairconf{\ctxone}{\mconfone}$, gives the 
same result of evaluating the \emph{monadic computation} 
$\ctxone[\mconfone] \in \monad (\computations)$, 
where for
$\mconfone = \eff{\gop}{n}{i}{\econfone_i}$, we define 
$\ctxone[\mconfone]$ pointwise as $\eff{\gop}{n}{i}{\ctxone[\econfone_i]}$.
Such a dynamics explicitly separates the computational 
steps acting on $\ctxone$ only from those making $\ctxone$ and $\mconfone$ interact. 
This feature is crucial, as it shows that any interaction between $\ctxone$ and 
$\mconfone$ corresponds 
to a $\strongkleisli{\syskappa}$-action, so that equivalent 
$\strongkleisli{\syskappa}$-states will have the same $\sysf$-dynamics when paired 
with the same context. 
That gives us a finer analysis of the computational behaviour of 
the compound monadic computation $\ctxone[\mconfone]$, and ultimately of a 
compound computation $\ctxone[\termone]$. 
As we did for $\syskappa$, it is actually convenient to determinise $\sysf$. 
We call the resulting system $\strongkleisli{\sysf}$.
Finally, from $\strongkleisli{\sysf}$ we can come back 
to $\monad(\Lambda)$ 
using the map $\push: \strongkleisli{\sysf} \to \monad(\Lambda)$ 
defined by 
$\push \left(\eff{\gop}{n}{i}{\pairconf{\ctxone}{\kappa}}\right) 
= \eff{\gop}{n}{i}{\ctxone[\kappa]}$.
We summarize the systems introduce in the following table.

\begin{center}
\begin{tabular}{|ccccc|}
\hline
\textbf{System} & $\syskappa$ & $\strongkleisli{\syskappa}$ & $\sysf$ & $\strongkleisli{\sysf}$
\\
\textbf{States} 
& Configurations $K$ 
& Monadic configurations $\kappa$
& Pairs $\pairconf{\ctxone}{\mconfone}$ 
& Monadic pairs
\\
\textbf{Dynamics} 
& Definition~\ref{def:system-k}
& Kleisli lifting of $\syskappa$
& $\strongkleisli{\sem{\ctxone[\mconfone]}}$ 
& Kleisli lifting of $\sysf$
\\
\hline
\end{tabular}
\end{center}

What remains to be clarified is how relations between computations 
can be transformed into relations on the aforementioned systems. The answer to 
this question is given by the following \emph{lax}\footnote{
  Each square gives a set-theoretic inclusion. For instance, the leftmost 
  square states that ${\lambdatraceeq} \subseteq {\kappatraceeq}$.
} commutative diagram:
$$
\xymatrixcolsep{3pc}
\xymatrix{
  \syslambda \ar@{^{(}->}[r] \ar[d]_{\lambdatraceeq}|-*=0@{|}
  & \syskappa \ar@{^{(}->}[r] \ar[d]_{\kappatraceeq}|-*=0@{|}
  & \strongkleisli{\syskappa} \ar[r]^{\ctxone[-]} 
    \ar[d]_{\kappastartraceeq}|-*=0@{|}
  & \sysf \ar@{^{(}->}[r]
    \ar[d]_{\mathtt{C}(\kappastartraceeq)}|-*=0@{|}
  & \strongkleisli{\sysf} \ar[r]^{\obsfstar} 
    \ar[d]_{\bc{\kappastartraceeq}}|-*=0@{|}
  & \monad 1 \ar[d]_{=}|-*=0@{|}
  \\
  \syslambda \ar@{^{(}->}[r]
   & \syskappa \ar@{^{(}->}[r]
  & \strongkleisli{\syskappa} \ar[r]_{\ctxone[-]} 
  & \sysf \ar@{^{(}->}[r]
  & \strongkleisli{\sysf} \ar[r]_{\obsfstar} 
  & \monad 1
} 
$$
Here, $\mathtt{C}(\relone)$ denotes the contextual closure of $\relone$, whereas 
$\mathtt{B}(\relone)$ is the Barr extension 
of $\relone$ \cite{Barr/LMM/1970,Kurz/Tutorial-relation-lifting/2016}. 
Finally, the map $\obsfstar$ is obtained postcomposing the observation map $\obs$ 
with $\push$.

\subsection{Determinisation: 
From $\syskappa$ to $\strongkleisli{\syskappa}$} 

The first step of our strategy is the 
determinisation of $\syskappa$.
We do so by taking advantage of Remark~\ref{rem:well-typed-configurations} 
and working with a transition system whose states are monadic configurations 
in $\bigcup_{\typeconfone} \monad(\envconf_{\typeconfone})$. Without much of a surprise,
we extend the notion of a type to monadic configurations by stipulating 
$\mconfone$ has type $\typeconfone$ if and only if 
$\mconfone \in \monad(\envconf_{\typeconfone})$.

\begin{definition}
\label{def:kappa-star}
System $\strongkleisli{\syskappa}$ has 
elements in $\bigcup_{\typeconfone} \monad(\envconf_{\typeconfone})$ 
as states, $\syskappa$-actions as actions, and transition structure thus defined, 
where $\eff{\gop}{n}{i}{\econfone_i} \in \monad(\envconf_{\typeconfone})$:
\[
\infer{
\eff{\gop}{n}{i}{\econfone_i} 
\xRightarrow{\ \ell\ }
\eff{\gop}{n}{i}{\mconfone_i}
}
{\typefun(\typeconfone, \ell) \textnormal{ defined } 
&
\forall i \in n.\ \econfone_i
\xrightarrow{\ \ell\ } \mconfone_i}
\]
\end{definition}

Notice that $\strongkleisli{\syskappa}$ is indeed a deterministic system
and that, by Remark~\ref{rem:well-typed-configurations}, the transition 
structure of $\strongkleisli{\syskappa}$ is well-defined. For suppose 
$\mconfone = \eff{\gop}{n}{i}{\econfone_i}$ is a $\strongkleisli{\syskappa}$-state, and 
thus an element in $\monad(\envconf_{\typeconfone})$ for some configuration 
type $\typeconfone$, and let $\ell$ be an action such that 
$\typefun(\typeconfone,\ell) = \typeconftwo$.
Then, for any 
$i \in n$, we have
$\econfone_i \xrightarrow{\ \ell\ } \mconfone_i$, for some 
$\mconfone_i \in \monad(\envconf_{\typeconftwo})$. As a consequence, we see that
$\eff{\gop}{n}{i}{\mconfone_i} \in \monad(\envconf_{\typeconftwo})$.
\longv{
Notice also that for $\boteffn{\typeconfone} \in \monad(\envconf_{\typeconfone})$, we have 
$\boteffn{\typeconfone} \xRightarrow{\ a\ } \boteffn{\typeconftwo}$, for any 
$\typeconftwo$ and action $a$ such that $\typefun(\typeconfone,a) = \typeconftwo$.
}

We define a notion of trace equivalence 
for $\strongkleisli{\syskappa}$ pretty much as we did for 
$\syskappa$. The action sets of $\syskappa$ and 
$\strongkleisli{\syskappa}$ being the same, the set of 
traces of $\syskappa$ and 
$\strongkleisli{\syskappa}$ are the same as well. Moreover, given 
a $\strongkleisli{\syskappa}$-state $\mconfone$, the set 
$Tr(\mconfone)$ is defined in the obvious way. 
Finally, we rely on the structure $(\monad\{*\}, \obskappastar)$ for observations, 
where $\obskappastar$ maps a $\strongkleisli{\syskappa}$-state to an element in 
$\monad \{*\}$ as usual: $
\obskappastar \Big(\eff{\gop}{n}{i}{\econfone_i}\Big)
\defeq \eff{\gop}{n}{i}{*}$.

\longv{
  \begin{definition}
  Let $\mconfone \in \monad(\envconf_{\typeconfone})$ and 
  $\traceone \in Tr(\mconfone)$. Define the element 
  $\strongkleisli{\sat}(\mconfone, \traceone) \in \monad 1$ as follows:
  \begin{align*} 
  \strongkleisli{\sat}(\mconfone, \varepsilon) 
  &\defeq \obskappastar(\mconfone);
  &
  \strongkleisli{\sat}(\mconfone, a \cdot \tracetwo) 
  &\defeq \strongkleisli{\sat}(\mconftwo, \tracetwo) 
  \text{ where } \mconfone \xRightarrow{\ a\ } \mconftwo
  \end{align*}
  \end{definition}
}

\shortv{
  \begin{definition}
Let $\mconfone \in \monad(\envconf_{\typeconfone})$ and 
$\traceone \in Tr(\mconfone)$. Define 
$\strongkleisli{\sat}(\mconfone, \traceone) \in \monad\{*\}$ by stipulating that
$\strongkleisli{\sat}(\mconfone, \varepsilon) \defeq \obskappastar(\mconfone)$
and 
$\strongkleisli{\sat}(\mconfone, \ell \cdot \tracetwo) 
\defeq \strongkleisli{\sat}(\mconftwo, \tracetwo)$, with
$\mconfone \xRightarrow{\ \ell\ } \mconftwo$.
\end{definition}
}

\emph{Trace equivalence} $\kappastartraceeq$ is the relation on 
$\strongkleisli{\syskappa}$-states thus defined:
$$
\mconfone \kappastartraceeq \mconftwo \iff
Tr(\mconfone) = Tr(\mconftwo) 
\wedge
\forall \traceone \in Tr(\mconfone).\ \strongkleisli{\sat}(\mconfone,\traceone) = 
\strongkleisli{\sat}(\mconftwo, \traceone).
$$

\shortv{
\begin{proposition}
For any $\strongkleisli{\syskappa}$-state $\mstateone = 
\eff{\gop}{n}{i}{\econfone_i}$ 
and trace $\traceone \in Tr(\mconfone)$, we have
$$
\strongkleisli{\sat} \Big(\eff{\gop}{n}{i}{\econfone_i},\traceone \Big) 
= \eff{\gop}{n}{i}{\sat(\econfone_i,\traceone)}
.$$ 
Therefore, for all $\syskappa$-states $\econfone, \econftwo$, we have  
$\econfone \kappatraceeq \econftwo$ if and only if 
$\effunit{\econfone} \kappastartraceeq \effunit{\econftwo}$.
\end{proposition}

\begin{proof}
First, observe that if $\traceone \in Tr(\mstateone)$, then $\traceone \in Tr(\econfone_i)$, 
for any $i$. In fact, say $\mstateone \in \monad(\envconf_{\typeconfone})$, 
so that $\econfone_i \in \envconf_{\typeconfone}$ for any $i$. 
Since whether $\traceone \in Tr(\econfone_i)$ is determined by the type of $\econfone_i$, 
we indeed have $\traceone \in Tr(\econfone_i)$, for any $i \in n$. 
We now prove $
\strongkleisli{\sat} \Big(\eff{\gop}{n}{i}{\econfone_i},\traceone \Big) 
= \eff{\gop}{n}{i}{\sat(\econfone_i,\traceone)}
$
by induction on $\traceone$. 
The case for $\traceone = \varepsilon$ is trivial. Suppose 
$\traceone = \actone \cdot \tracetwo$. Since $\traceone \in Tr(\mstateone)$, we have 
$\eff{\gop}{n}{i}{\econfone_i} \xRightarrow{\ \actone\ } 
\efftwo{\gop}{n}{i}{\goptwo_i}{m_i}{j}{\econftwo_j}$, with 
$\econfone_i \xrightarrow{\ \actone\ } \eff{\goptwo_i}{m_i}{j}{\econftwo_j}$. 
We can thus compute:
\begin{align*}
\strongkleisli{\sat} \Big(\eff{\gop}{n}{i}{\econfone_i}, \actone \cdot \tracetwo \Big)
&= \strongkleisli{\sat} \Big(\efftwo{\gop}{n}{i}{\goptwo_i}{m_i}{j}{\econftwo_j}, \tracetwo \Big)
\\
&\stackrel{IH}{=} \efftwo{\gop}{n}{i}{\goptwo_i}{m_i}{j}{\sat(\econftwo_j,\tracetwo)}
\\
&= \eff{\gop}{n}{i}{\sat(\econfone_i, \actone \cdot \tracetwo)}
\end{align*}
\end{proof}

Finally, we take advantage of the deterministic nature of $\strongkleisli{\syskappa}$ 
and characterise trace equivalence coinductively as 
$\strongkleisli{\syskappa}$-bisimilarity.
\begin{definition}
\label{definition:kappa-star-similarity}
Define $\strongkleisli{\syskappa}$-bisimilarity 
$\kappastarbisimilarity$ as the union of all symmetric relation  $\relone$ 
on $\strongkleisli{\syskappa}$-states
such that:
\begin{varenumerate} 
\item $\mconfone \relone \mconftwo$ and $\mconfone \xRightarrow{\ \ell\ } \mconfone'$
imply $\mconftwo \xRightarrow{\ \ell\ } \mconftwo'$ and
  $\mconfone' \relone \mconftwo'$.
\item
  $\mconfone \relone \mconftwo$
  implies $\obskappastar(\mconfone) = \obskappastar(\mconftwo)$.
\end{varenumerate}
\end{definition}

\begin{proposition} 
\label{prop:similarity-coincides-with-traceleq-bang}
${\kappastarbisimilarity} = {\kappastartraceeq}$.
\end{proposition}

\begin{proof}
Obviously $\kappastarbisimilarity$ is contained in $\kappastartraceeq$. For the converse, 
observe that $\kappastartraceeq$ is a bisimulation. 
\end{proof}

}

\longv{
\begin{lemma}
Given a $\strongkleisli{\syskappa}$-state $\mstateone = \eff{\gop}{n}{i}{\econfone_i}$ 
and a trace $t \in Tr(\mstateone)$, we have:
$$
\strongkleisli{\sat} \Big(\eff{\gop}{n}{i}{\econfone_i},t \Big) 
= \eff{\gop}{n}{i}{\sat(\econfone_i,t)}
$$
\end{lemma}

\begin{proof}
First of all observe that if $t \in Tr(\mstateone)$, then $t \in Tr(\econfone_i)$, 
for any $i$. In fact, say $\mstateone \in \monad(\envconf_{\typeconfone})$, 
so that $\econfone_i \in \envconf_{\typeconfone}$ for any $i$. 
Since whether $t \in Tr(\econfone_i)$ is determined by the type of $\econfone_i$, 
we indeed have $t \in Tr(\econfone_i)$, for any $i \in n$. We now prove the thesis 
by induction on $t$. The case for $t = \varepsilon$ is trivial. Suppose 
$t = a \cdot u$. Since $t \in Tr(\mstateone)$, we have 
$\eff{\gop}{n}{i}{\econfone_i} \xRightarrow{\ a\ } 
\efftwo{\gop}{n}{i}{\goptwo_i}{m_i}{j}{\econftwo_j}$, with 
$\econfone_i \xrightarrow{\ a\ } \eff{\goptwo_i}{m_i}{j}{\econftwo_j}$. 
We can thus compute:
\begin{align*}
\strongkleisli{\sat} \Big(\eff{\gop}{n}{i}{\econfone_i}, a \cdot u \Big)
&= \strongkleisli{\sat} \Big(\efftwo{\gop}{n}{i}{\goptwo_i}{m_i}{j}{\econftwo_j}, u \Big)
\\
&\stackrel{IH}{=} \efftwo{\gop}{n}{i}{\goptwo_i}{m_i}{j}{\sat(\econftwo_j,u)}
\\
&= \eff{\gop}{n}{i}{\sat(\econfone_i, a \cdot u)}
\end{align*}
\end{proof}

\begin{corollary}
Given two $\syskappa$-states $\econfone, \econftwo$, we have 
$\econfone \kappatraceleq \econftwo$ if and only if 
$\effunit{\econfone} \kappastartraceleq \effunit{\econftwo}$.
\end{corollary}

Finally, we take advantage of the deterministic nature of $\strongkleisli{\syskappa}$ 
and characterise trace equivalence coinductively as 
$\strongkleisli{\syskappa}$-bisimilarity.

\begin{definition}
Define $\strongkleisli{\syskappa}$-bisimilarity 
$\kappastarbisimilarity$ as the largest relation $\relone$ 
on $\strongkleisli{\syskappa}$-states
such that:
\begin{itemize}
  \item
  $\mstateone \relone \mstatetwo$ and $\mstateone \xRightarrow{\ a\ } \mstateone'$
  implies $\mstatetwo \xRightarrow{\ a\ } \mstatetwo'$ and
  $\mstateone' \relone \mstatetwo'$
  \item
  $\mstateone \relone \mstatetwo$
  implies $\obskappastar(\mstateone) \cpoleq \obskappastar(\mstatetwo)$.
\end{itemize}
\end{definition}

As usual, since $\obskappastar$ is monotone we can define $\kappasimilarity$ 
coinductively as 
the greatest fixed point of a suitable monotone function. Moreover, 
$\strongkleisli{\syskappa}$ being deterministic, we can recover 
$\strongkleisli{\syskappa}$-bisimilarity as the intersection of 
$\kappasimilarity$ and its dual.

\begin{proposition} 
\label{prop:similarity-coincides-with-traceleq-bang}
${\kappasimilarity} = {\kappastartraceleq}$.
\end{proposition}

\begin{proof}
Obviously $\kappasimilarity$ is contained in $\kappastartraceleq$. For the converse, 
observe that $\kappastartraceleq$ is a simulation. 
\end{proof}

Finally, we recover standard inductive reasoning on finite approximations of 
program semantics by means of finite-step simulation. 

\begin{lemma}
  Let $k \geq 0$. Define system $\syskappa_k$ by replacing 
  $\seml{\termone}$ with $\semln{\termone}{k}$ in 
  Definition~\ref{def:system-kappa}, and system $\strongkleisli{\syskappa}_k$ 
  by replacing $\syskappa$ with $\syskappa_k$ in Definition~\ref{def:kappa-star} 
  Let $\kappasimilarity_{k}$ be similarity on $\strongkleisli{\syskappa}_k$ 
  and define \emph{finite-step similarity} $\finkappastarsimilarity$ as
  $\bigcap_k \kappasimilarity_{k}$. 
  Then ${\finkappastarsimilarity} = {\kappasimilarity}$.
\end{lemma}

\begin{proof}[Proof sketch]
The hard part is proving ${\finkappastarsimilarity} \subseteq {\kappasimilarity}$. 
For that we show that the $\strongkleisli{\syskappa}$-relation 
$\relone \defeq \{(\lub_n \mconfone_n, \mconftwo) \mid \mconfone_n 
\finkappastarsimilarity \mconftwo\}$ is a $\strongkleisli{\syskappa}$-simulation. 
To do so we rely on the $\ocppo$-enrichment of $\monad$ and use 
diagonalisation of double chains (given a sequence $(x_{n,m})_{n,m}$ 
in a domain, if $n \leq n'$ and $m \leq m'$ imply $x_{n,m} \cpoleq x_{n',m'}$, 
then $\lub_n \lub_m x_{n,m} = \lub_m \lub_n x_{n,m} = \lub_k x_{k,k}$).
\end{proof}
}

\subsection{From $\strongkleisli{\syskappa}$ to $\strongkleisli{\sysf}$}

The next step of our construction is to equip $\strongkleisli{\syskappa}$-states 
with contexts. To do so, we first define the notion of a \emph{context for a
configuration}. Without much of a surprise, the latter is modelled as an open term $t$
whose free variables can be instantiated with terms in configurations. 
However, in order to properly account for configurations of the form 
$\econfterm{\gone}{\done}{\termone}$, we have to consider open terms 
having a free variable (one is enough for our purposes) acting as a 
placeholder for a \emph{linearly-used} computations. 

\longv{
  \begin{definition} 
  Let $\econf{\gone}{\done}$ be a configuration with $|\gone| = n$ and 
  $|\done| = m$. 
  A context for $\econf{\gone}{\done}$ is simply a $(n,m)$-context, i.e. an open term 
  $\env{\evarone_1, \hh, \evarone_n}{\varone_1, \hh, \varone_m}
  \imp t$.
  A context for a configuration
  $\econfterm{\gone}{\done}{\termone}$ is an open term 
  $\envterm{\evarone_1, \hh, \evarone_n}{\varone_1, \hh, \varone_m}{\cvarone}
  \imp t$, where $\cvarone$ is a \emph{linear} placeholder for a computation. 
  \end{definition}
}
\shortv{
\begin{definition} 
Let $\econf{\gone}{\done}$ be a configuration with $|\gone| = n$ and 
$|\done| = m$. 
A context for $\econf{\gone}{\done}$ is simply a $(n,m)$-context, 
while a context for
$\econfterm{\gone}{\done}{\termone}$ is an open term 
$\envterm{\evarone_1, \hh, \evarone_n}{\varone_1, \hh, \varone_m}{\cvarone}
\imp t$, where $\cvarone$ is a \emph{linear} placeholder for a computation. 
\end{definition}
}

Due to space constraints, we do not given an explicit system 
for sequents of the form $\envterm{\eenvone}{\lenvone}{\cvarone} \imp t$, 
as such a system is standard.

Given a monadic configurations $\mconfone$, we say that 
$t$ is a context for $\mconfone$ if $t$ is a context for all
configurations in $\support{\mconfone}$ (notice that  
if $t$ is a context for a configuration in the support of $\mconfone$, 
then it is a context for all such configurations). 
If that is the case, then we can pair 
$t$ and $\mconfone$ together, obtaining a 
monadic term $t[\mconfone] \in \monad (\computations) \cup \monad (\values)$, where for 
$\mconfone = \eff{\gop}{n}{i}{\econfone_i}$ we define $t[\mconfone]$ as
$
\eff{\gop}{n}{i}{t[\econfone_i]}
$.

\renewcommand{\ctxone}{t}
\renewcommand{\ctxtwo}{s}

In order to study the computational behaviour of a $\strongkleisli{\syskappa}$-state
paired with a context for it, we
define a new
system, called $\sysf$, whose states are figures of the form 
$\pairconf{t}{\mconfone}$, with $t$ context for $\mconfone$. 
If $t[\mconfone] \in \monad( \values)$, then we say that $\pairconf{\ctxone}{\mconfone}$ 
is a $\sysf$-value state 
(similarity, we have $\sysf$-computation states when 
$t[\mconfone] \in \monad (\computations)$: by type assumption, these are 
the only possible cases).
The dynamics of $\sysf$ is given by an evaluation function $\semf{-}$
mapping $\sysf$-computation states to monadic $\sysf$-value states.

In order to facilitate the definition of $\semf{-}$, it is 
convenient to first extend the action of $\seml{-}$ to \emph{open} terms. 
We do so following \cite{Lassen/EagerNormalFormBisimulation/2005}.

\longv{
  \begin{definition}
    An evaluation context $\ectxone$ is an expression generated by the following 
    grammar: 
    \begin{align*}
      \ectxone &::= [-] \mid \seq{\ectxone}{\termone}.
    \end{align*}
    A stuck expression is an expression of the form $\ectxone[\stuckterm]$, 
    where $\stuckterm$ is an expression having the shape of a redex but whose 
    evaluation is stuck, due to the presence of a variable. Formally, 
    stuck expressions are thus defined:
    \begin{align*}
    \stuckterm &\bnf \evarone \mid \cvarone \mid \varone\valone \mid 
      \coseq{\varone}{\termone}
    \end{align*} 
  \end{definition}
}

\shortv{
\begin{definition}
  A \emph{stuck term} is a term of the form 
  $\ectxone[\stuckterm]$, where 
  $\stuckterm$ is thus defined:
  \begin{align*}
   \stuckterm &\bnf \evarone \mid \cvarone \mid \varone\valone \mid 
    \coseq{\varone}{\termone}
  \end{align*}
\end{definition}
}

We are now ready to define the dynamics $\semf{-}$ of system
$\sysf$. As usual, we define $\semf{-}$ as $\lub_{k \geq 0} \semfn{-}{k}$, 
where  
$\semfn{\pairconf{\ctxone}{\mconfone}}{0} \defeq \boteff$. 
In order to define 
$\semfn{\pairconf{\ctxone}{\mconfone}}{k}$, where $\pairconf{\ctxone}{\mconfone}$ 
is a $\sysf$-computation state, we proceed by cases on 
$\ctxone$.
\longv{
\begin{description}
  \item[Case 1.] Suppose $\ctxone$ is neither a value nor a stuck term.  
    Then $\semfn{\pairconf{\ctxone}{\mstateone}}{k+1}$ 
    simply evaluates $\ctxone$.
    \begin{align*}
      \semfn{\pairconf{\val{\ctxone}}{\mstateone}}{k+1} 
      &\defeq \effunit{\pairconf{\ctxone}{\mstateone}}
      \\
      \semfn{\pairconf{\ectxone[\seq{(\val \ctxone)}{\ctxtwo}]}{\mstateone}}{k+1} 
      &\defeq \semfn{\pairconf{\ectxone[\subst{\ctxtwo}{\varone}{\ctxone}]}
      {\mstateone}}{k}
      \\
      \semfn{\pairconf{\ectxone[(\abs{\varone}{\ctxtwo})\ctxone]}{\mstateone}}{k+1}
      &\defeq \semfn{\pairconf{\ectxone[\subst{\ctxtwo}{\varone}{\ctxone}]}{\mstateone}}{k}
      \\ 
      \semfn{\pairconf{\ectxone[\coseq{\bang{\ctxone}}{\ctxtwo}]}{\mstateone}}{k+1}
      &\defeq \semfn{\pairconf{\ectxone[\subst{\ctxtwo}{\evarone}{\ctxone}]}{\mstateone}}
      {k}
      \\
      \semfn{\ectxone[\op(\ctxone_1, \hh, \ctxone_n)]}{k+1}
      &\defeq \eff{\sem{\op}}{n}{i}{\semfn{\ectxone[\ctxone_i]}{k}}
      \end{align*}
    \item[Case 2.] Suppose $\ctxone$ is of the form $\ectxone[\cvarone]$.
      We do a further analysis on the shape of $\ectxone$. 
      In the following, we write $\semkn{\econfone}{k}$, where 
      $\econfone$ is a configuration of the form 
      $\econfterm{\gone}{\done}{\termone}$, for 
      $\eff{\gop}{n}{i}{\econf{\gone}{\done, \valone_i}}$, where 
      $\semln{\termone}{k} = \eff{\gop}{n}{i}{\valone_i}$. We extend 
      $\semk{-}$ to a map $\semkstar{-}$ acting on $\strongkleisli{\syskappa}$-states 
      as usual. 
      \begin{description}
        \item[Case 2.1.] 
          Consider the case for $\pairconf{\cvarone}{\mstateone}$. Since 
          $\cvarone$ is a context for $\mstateone$, $\mstateone$ must have 
          the form $\eff{\gop}{n}{i}{\econfterm{\emptyenv}{\emptyenv}{\termone_i}}$,
          so that $\semkstarn{\mstateone}{k}$ have the form 
          $\eff{\goptwo}{m}{j}{\econf{\emptyenv}{\valone_j}}$. Define:
          $$
          \semfn{\pairconf{\cvarone}{\mstateone}}{k+1} 
          \defeq \effunit{\pairconf{\varone_1}{\semkstarn{\mstateone}{k}}}.
          $$
        \item[Case 2.2.] Consider the case for 
          $\pairconf{\ectxone[\seq{\cvarone}{\ctxtwo}]}{\mstateone}$. As before, 
          we must have that any configuration $\econfone$ in the support of
          $\mstateone$ must have the form $\econfterm{\gone}{\done}{\termone}$. 
          Therefore, any configuration in the support of $\semkstarn{\econfone}{k}$ 
          must have the form $\econf{\gone}{\done, \valone}$. 
          Let $|\done| = n$. Define:
          $$
          \semfn{\pairconf{\ectxone[\seq{\cvarone}{\ctxtwo}]}{\mstateone}}{k+1}
          \defeq 
          \semfn{\pairconf{\ectxone[\subst{\ctxtwo}{\varone}{\varone_{n+1}}]}
          {\semkstarn{\mstateone}{k}}}{k}
          $$
      \end{description}
    \item[Case 3.] We have to consider those cases $\ectxone[\stuckterm]$ where the stuck 
      expression $\stuckterm$ comes from a variable acting as a placeholder 
      for a resource in a configuration of the form $\econf{\gone}{\done}$. 
      In those cases we just mimic transitions in 
      $\strongkleisli{\syskappa}$ and update $\ectxone[\stuckterm]$ accordingly. 
      Notice that 
      this is exactly what we have done in case 2, where we have 
      mimicked $\eval$ actions. 
      \begin{itemize}
        \item
          Consider the case for $\ectxone[\evarone_i]$. Since the latter 
          is a context for $\mstateone$, any configuration $\econfone$ in the support 
          of $\mstateone$ must have the form $\econf{\gone[\termone]_i}{\done}$. 
          As a consequence, we have the $\syskappa$-transition 
          $\econfone \xrightarrow{\ !_i\ } 
          \effunit{\econfterm{\gone[\termone]_i}{\done}{\termone}}$, and thus 
          a $\strongkleisli{\syskappa}$-transition from $\mstateone$, say to
          $\mstatetwo$. Define:
          $$
          \semfn{\pairconf{\ectxone[\evarone_i]}{\mstateone}}{k+1}
          \defeq \semfn{\pairconf{\ectxone[\cvarone]}{\mstatetwo}}{k}
          $$
        \item Consider the case for $\ectxone[\coseq{\varone_i}{\ctxone}]$. 
          Since the latter 
          is a context for $\mstateone$, any configuration $\econfone$ in the support 
          of $\mstateone$ must have the form $\econf{\gone}{\done[\bang{\termone}]_i}$. 
          As a consequence, we have the $\syskappa$-transition 
          $\econfone \xrightarrow{\ ?_i\ } 
          \effunit{\econf{\gone, \termone}{\done}}$, and thus 
          a $\strongkleisli{\syskappa}$-transition from $\mstateone$, say to
          $\mstatetwo$. Let $|\gone| = n$. Define:
          $$
          \semfn{\pairconf{\ectxone[\coseq{\varone_i}{\ctxone}]}{\mstateone}}{k+1}
          \defeq \semfn{\pairconf{\ectxone[\coseq{\bang{\evarone}_{n+1}}{\ctxone}]}
          {\mstatetwo}}{k}.
          $$
        \item Finally, consider the case for $\ectxone[\varone_i \ctxone]$.
          Since the latter is a valid context for $\mstateone$, 
          (i) any configuration 
          $\econfone$ in the support of $\mstateone$ must have the form
          $\econf{\gone}{\done[\abs{\varone}{\termtwo}]_i}$, (ii) 
          there must 
          exist sequences $i \not \in \bar{s}, \bar{p}$ such that
          $\bar{s}, \bar{p}$ are valid for $\gone, \done$ respectively, 
          and (iii) $\ctxone$ is a $(|\bar{s}|, |\bar{p}|)$-value context (and thus 
          $\ctxone$ open value). 
          As a consequence, we have the $\syskappa$-transition
          $$
          \econf{\gone}{\done[\abs{\varone}{\termtwo}]_i} 
          \xrightarrow{\ (\bar{s}, \bar{p}, i, \ctxone)\ }
          \effunit{\econfterm{\gone}{\done \ominus \bar{p}}{(\abs{\varone}{\termtwo}) 
          \ctxone[\gone_{\bar{s}}, \done_{\bar{p}}]}}
          $$
          and thus $\mstateone \xRightarrow{\ (\bar{s}, \bar{p}, i, \ctxone)\ } \mstatetwo$, 
          for a suitable $\strongkleisli{\syskappa}$-state $\mstatetwo$. Let 
          $|\done \ominus \bar{p}| = n$. Define:
          $$
          \semfn{\pairconf{\ectxone[\varone_i \ctxone]}{\mstateone}}{k+1}
          \defeq \semfn{\pairconf{\ectxone^{\star}[\cvarone]}{\mstatetwo}}{k}
          $$
          where $\ectxone^{\star}$ is the re-indexing of free variables of $\ectxone$
          according to $\done \ominus \bar{p}$. That is, recall that a context 
          for $\econf{\gone}{\done}$ with $|\gone| =n$, $|\done| = m$, 
          is a term $\env{\evarone_1, \hh, \evarone_n}{\varone_1, \hh, \varone_m} 
          \imp \ctxone$ with the intended meaning that, e.g., variable 
          $\varone_i$ is a placeholder for the $i$-th value in $\done$. Say 
          the latter is $\valone$. 
          When passing from $\done$ to $\done \ominus \bar{p}$ we change the
          position of values in $\done$, so that the $i$-th value in $\done$ (i.e. $\valone$)
          (to which we associate the variable $\varone_i$ in $\ctxone$) may 
          not be at position $i$ in $\done \ominus \bar{p}$. Therefore, we have 
          to change the index $i$ in $\varone_i$ to an index $j$ in such a 
          way that $\varone_j$ is associated to $\valone$ in $\done \ominus \bar{p}$. 
          Such a re-indexing can be easily done observing that if 
          $\valone$ has position $i$ in $\done$, then it has position 
          $i - |\{p \in \bar{p} \mid p < i\}|$ in $\done \ominus \bar{p}$. 
      \end{itemize}
\end{description}
}
\shortv{
Obviously, $\ctxone$ cannot be a value. If $\ctxone$ 
is not a stuck term,
then $\semfn{\pairconf{\ctxone}{\mconfone}}{k}$ 
simply evaluates $\ctxone$ according to the first three 
rules in Figure~\ref{fig:operational-semantics-lambda-bang-open}.
Otherwise, $\ctxone$ is a term of the form $\ectxone[\stuckterm]$.
We proceed by mimicking transitions of 
$\strongkleisli{\syskappa}$ and updating (as well as adding) variables in 
$\ectxone[\stuckterm]$ accordingly. Due to space constraints, 
we examine only a couple of relevant cases.
Let us write $\semk{\econfone}$, where 
$\econfone$ is a configuration of the form 
$\econfterm{\gone}{\done}{\termone}$, for 
$\eff{\gop}{n}{i}{\econf{\gone}{\done, \valone_i}}$, with 
$\seml{\termone} = \eff{\gop}{n}{i}{\valone_i}$. We extend 
$\semk{-}$ to a map $\semkstar{-}$ acting on $\strongkleisli{\syskappa}$-states 
pointwise. 
\begin{varitemize}
\item 
Consider the case for $\pairconf{\cvarone}{\mconfone}$. Since 
          $\cvarone$ is a context for $\mconfone$, $\mconfone$ must have 
          the form $\eff{\gop}{n}{i}{\econfterm{\emptyenv}{\emptyenv}{\termone_i}}$,
          so that $\semkstar{\mconfone}$ have the form 
          $\eff{\goptwo}{m}{j}{\econf{\emptyenv}{\valone_j}}$. Define
          $
          \semfn{\pairconf{\cvarone}{\mconfone}}{k+1}
          $
          as
          $ 
          \effunit{\pairconf{\varone_1}{\semkstar{\mconfone}}}.
          $
\item
Consider now the case for $\ectxone[\varone_l \ctxtwo]$.
          Since the latter is a valid context for $\mconfone$, 
          (i) any configuration 
          $\econfone \in \support{\mconfone}$ must have the form
          $\econf{\gone}{\done[\abs{\varone}{\termtwo}]_l}$, (ii) 
          there must 
          exist sequences $l \not \in \bar{\imath}, \bar{\jmath}$ such that
          $\bar{\imath}, \bar{\jmath}$ are valid for $\gone, \done$, respectively, 
          and (iii) $\ctxtwo$ is a $(|\bar{\imath}|, |\bar{\jmath}|)$-value context. 
          As a consequence, we have the $\syskappa$-transition
          $$
          \econf{\gone}{\done[\abs{\varone}{\termtwo}]_l} 
          \xrightarrow{\ (\bar{\imath}, \bar{\jmath}, l, \ctxtwo)\ }
          \effunit{\econfterm{\gone}{\done \ominus \bar{\jmath}}
          {\subst{\termtwo}{\varone}{\ctxtwo[\gone_{\bar{\imath}}, \done_{\bar{\jmath}}]}}}
          $$
          and thus $\mconfone \xRightarrow{\ (\bar{\imath}, \bar{\jmath}, j, \ctxtwo)\ } 
          \mconftwo$, 
          for a suitable $\strongkleisli{\syskappa}$-state $\mconftwo$. 
          Define
          $
          \semfn{\pairconf{\ectxone[\varone_l \ctxtwo]}{\mconftwo}}{k+1}
          $ as
          $\semfn{\pairconf{\ectxone^{\star}[\cvarone]}{\mconftwo}}{k}
          $,
          where $\ectxone^{\star}$ is the re-indexing of free variables of $\ectxone$
          according to $\done \ominus \bar{\jmath}$. That is, recall that a context 
          for $\econf{\gone}{\done}$ with $|\gone| =n$, $|\done| = m$, 
          is a term $\env{\evarone_1, \hh, \evarone_n}{\varone_1, \hh, \varone_m} 
          \imp \ctxone$ with the intended meaning that, e.g., variable 
          $\varone_i$ is a placeholder for the $i$-th value in $\done$. Say 
          the latter is $\valone$. 
          When passing from $\done$ to $\done \ominus \bar{c}$, 
          for a (suitable) sequence $c$, we change the
          position of values in $\done$, so that the $i$-th value in $\done$ (i.e. $\valone$)
          (to which we associate the variable $\varone_i$ in $\ctxone$) may 
          not be at position $i$ in $\done \ominus \bar{c}$. Therefore, we have 
          to change the index $i$ in $\varone_i$ to an index $j$ in such a 
          way that $\varone_j$ is associated to $\valone$ in $\done \ominus \bar{c}$. 
          Such a re-indexing can be easily done observing that if 
          $\valone$ has position $i$ in $\done$, then it has position 
          $i - |\{c \in \bar{c} \mid c < i\}|$ in $\done \ominus \bar{c}$. 
 \end{varitemize}
 \noindent 
 The complete list of the defining rules of 
 $\semfn{\pairconf{\ctxone}{\mconfone}}{k+1}$ is summarised
in Figure~\ref{fig:operational-semantics-lambda-bang-open}, 
where we employ the notation used in the above discussion.
}

\longv{
\noindent We summarise the defining rules of $\semfn{\pairconf{\ctxone}{\mconfone}}{k+1}$ 
in Figure~\ref{fig:operational-semantics-lambda-bang-open}, 
where we employ the notation used in the above discussion. 
}

\begin{figure*}[htbp]
\hrule 

\vspace{0.2cm}
\begin{align*}
\semfn{\pairconf{\val{\ctxone}}{\mconfone}}{k+1} 
      &\defeq \effunit{\pairconf{\ctxone}{\mconfone}}
      \\
       \semfn{\pairconf{\ectxone[\redex]}{\mconfone}}{k+1} 
      &\defeq \semfn{\pairconf{\ectxone[\redex']}
      {\mconfone}}{k}
      \\
      \semfn{\pairconf{\ectxone[\op(\ctxone_1, \hh, \ctxone_n)]}{\mconfone}}{k+1}
      &\defeq \eff{\sem{\op}}{n}{i}{\semfn{\pairconf{\ectxone[\ctxone_i]}{\mconfone}}{k}}
      \\
      \semfn{\pairconf{\cvarone}{\mconfone}}{k+1} 
      &\defeq \effunit{\pairconf{\varone_1}{\semkstar{\mconfone}}}.
      \\
      \semfn{\pairconf{\ectxone[\seq{\cvarone}{\ctxtwo}]}{\mconfone}}{k+1}
      &\defeq 
      \semfn{\pairconf{\ectxone[\subst{\ctxtwo}{\varone}{\varone_{n+1}}]}
      {\semkstar{\mconfone}}}{k}
      \\
      \semfn{\pairconf{\ectxone[\evarone_i]}{\mconfone}}{k+1}
      &\defeq \semfn{\pairconf{\ectxone[\cvarone]}{\mconftwo}}{k}.
      \\
      \semfn{\pairconf{\ectxone[\coseq{\varone_i}{\ctxtwo}]}{\mconfone}}{k+1}
      &\defeq \semfn{\pairconf{\ectxone[\coseq{\bang{\evarone}_{n+1}}{\ctxtwo}]}
      {\mconftwo}}{k}.
      \\
      \semfn{\pairconf{\ectxone[\varone_i \ctxone]}{\mconfone}}{k+1}
      &\defeq \semfn{\pairconf{\ectxone^{\star}[\cvarone]}{\mconftwo}}{k}
      \end{align*}
\vspace{-0.2cm}
\hrule
\caption{Definition of $\semfn{-}{k+1}$}
\label{fig:operational-semantics-lambda-bang-open}
\end{figure*}

Finally, we determinise $\sysf$ building a new system, which we call 
$\strongkleisli{\sysf}$. 

\begin{definition} 
System $\strongkleisli{\sysf}$ has monadic (well-typed) $\sysf$-states 
as states, 
where, as usual, all $\sysf$-states in the support of 
a $\strongkleisli{\sysf}$-state $\mfstarone$ have the same type. 
Given a $\strongkleisli{\sysf}$-state $\mfstarone$, if 
all  $\sysf$-states in its support are  $\sysf$-value states, then 
we say that $\mfstarone$ is a $\strongkleisli{\sysf}$-value state 
(and similarity for computation-states).
The dynamics of $\strongkleisli{\sysf}$ is given by the map $\semfstar{-}$ mapping 
$\strongkleisli{\sysf}$-computation states to $\strongkleisli{\sysf}$-value states:
$
\Bigl \llbracket 
\eff{\gop}{n}{i}{\pairconf{\ctxone_i}{\mconfone_i}}
\Bigr \rrbracket^{\scriptscriptstyle{\strongkleisli{\sysf}}} 
\defeq \eff{\gop}{n}{i}{\semf{\pairconf{\ctxone_i}{\mconfone_i}}}.
$
The map $\semfstar{-}_k$ is defined as usual.
\end{definition}

We can extract an element in $\monad( \values)$ 
out of a $\strongkleisli{\sysf}$-value state
(and an element in $\monad (\computations)$ out of a 
$\strongkleisli{\sysf}$-computation state)
using the function $\push$ mapping 
a $\strongkleisli{\syskappa}$-state
$\eff{\gop}{n}{i}{\pairconf{\ctxone_i}{\mconfone_i}} 
$ to $\eff{\gop}{n}{i}{\ctxone_i[\mconfone_i]}.
$
As expected, $\push$ connects 
$\semfstar{-}$ and $\semlstar{-}$.

\begin{lemma}
\label{lemma:push-eval}
For any
$\strongkleisli{\sysf}$-computation state $\mfstarone$, we have 
$
\push\ \semfstar{\mfstarone} = \semlstar{\push\ \mfstarone}
$.
\end{lemma}

\begin{proof}[Proof sketch]
This essentially follows from the way we have defined $\semf{-}$. 
In fact, it is sufficient to prove that for any $k \geq 0$, 
and $\sysf$-state $\pairconf{\ctxone}{\mconfone}$ we have:
\begin{align*}
\push\ \semfn{\pairconf{\ctxone}{\mconfone}}{k} &\cpoleq 
\sem{\ctxone[\mconfone]}^{\Lambda^{*}}
\\
\semn{\ctxone[\mconfone]}{k}^{\Lambda^{*}} &\cpoleq 
\push\ \semf{\pairconf{\ctxone}{\mconfone}}
\end{align*}
The proof proceeds by induction on $k$.
\end{proof}

We have thus came up with a way to relate system $\strongkleisli{\sysf}$ 
with monadic terms. We summarise such a relationship in the 
following commutative diagram, where $\obsfstar$ abbreviates 
$\obslambdastar \comp {\push}$, and
we write 
$\strongkleisli{\sysf}_{\computations}$ for the restriction of $\strongkleisli{\sysf}$ 
to $\strongkleisli{\sysf}$-computation states 
(similarity, we use the subscript $\values$ for 
$\strongkleisli{\sysf}$-value states).
$$
 \xymatrix@C=4pc{
   \monad \computations  
   \ar[r]^{\semlstar{-}}
   & \monad \values \ar[r]^{\obslambdastar}
    & \monad 1
  \\
  \strongkleisli{\sysf}_{\computations}
  \ar[r]^{\semfstar{-}}
  \ar[u]^{\push}
  & \strongkleisli{\sysf}_{\values} \ar[ru]_{\obsfstar}
 \ar[u]^{\push}
   & 
 }
$$

Notice that the $\ocppo$-enrichment of $\monad$ implies 
continuity of $\push$, and thus
of $\obsfstar$, since $\obslambdastar$ is continuous.

\renewcommand{\ctxone}{C}
We are now ready to prove soundness of $\lambdatraceeq$ for $\ctxeq$. 
Concretely, what we have to prove is that $\termone \lambdatraceeq \termtwo$ 
implies $\obslambdastar \seml{\ctxone[\termone]} = 
\obslambdastar \seml{\ctxone[\termtwo]}$, for any context $\ctxone$. 
To prove such a statement we need to study the computational 
behaviour of $\ctxone[\termone]$ and $\ctxone[\termtwo]$. The right setting to do so, 
is, obviously, system $\strongkleisli{\sysf}$. Hence, we need to move from 
programs to $\strongkleisli{\sysf}$-states. We do so 
by mapping $\ctxone[\termone]$ to 
$\mfstarone_{\termone} \defeq \effunit{
\pairconf{\ctxone}{\effunit{\econfterm{\emptyenv}{\emptyenv}{\termone}}}
}$ 
(ans similarity we map $\ctxone[\termtwo]$ to a $\strongkleisli{\sysf}$-state 
$\mfstarone_{\termtwo}$). By Lemma~\ref{lemma:push-eval}, 
we recover $\seml{\ctxone[\termone]}$ as 
$\push\ \semfstar{\mfstarone_{\termone}}$, and thus 
$\obslambdastar \seml{\ctxone[\termone]}$ as 
$\obsfstar \semfstar{\mfstarone_{\termone}}$. 
\longv{
  We thus find ourselves in a situation of the form:
  $$
  \xymatrix{
  \effunit{\Big(
  \pairconf{\ctxone}{\effunit{\econfterm{\emptyenv}{\emptyenv}{\termone}}}
  \Big)} 
  \ar[d] & \mathtt{F}({\lambdatraceleq}) & 
  \effunit{\Big(
  \pairconf{\ctxone}{\effunit{\econfterm{\emptyenv}{\emptyenv}{\termtwo}}}
  \Big)} \ar[d]
  \\
   \Big\llbracket \effunit{\Big(
  \pairconf{\ctxone}{\effunit{\econfterm{\emptyenv}{\emptyenv}{\termone}}}
  \Big)} \Big\rrbracket^{\strongkleisli{\sysf}}
  & ? 
  & \Big\llbracket \effunit{\Big(
  \pairconf{\ctxone}{\effunit{\econfterm{\emptyenv}{\emptyenv}{\termtwo}}}
  \Big)} \Big\rrbracket^{\strongkleisli{\sysf}} 
  }
  $$
  Here, $\mathtt{F}({\lambdatraceleq})$ is a lifting of 
  $\lambdatraceleq$ to $\strongkleisli{\sysf}$-states, and the question mark 
  $?$ stands for a relation on $\strongkleisli{\sysf}$-states such that:
  \begin{enumerate} 
    \item Whenever $\mstatethree \mathbin{\mathtt{F}({\lambdatraceleq})}  
      \mstatefour$, we have $\semfstar{\mstatethree} 
      \mathbin{?} 
      \semfstar{\mstatefour}$
    \item $\mstatethree \mathbin{?} \mstatefour$ implies 
    $\obsfstar(\mstatethree) \cpoleq \obsfstar(\mstatefour)$. 
  \end{enumerate}
  Obviously, if we are 
  able to find such a relation $?$ (as well as a suitable lifting $\mathtt{F}$), 
  we can conclude the wished thesis. 

  Let us examine how the definition of the definition of $\semfstar{-}$ proceeds. 
  First, we evaluate the context $\ctxone$, this way obtaining the diagram:
  $$
  \xymatrix{
  \effunit{\Big(
  \pairconf{\ctxone}{\effunit{\econfterm{\emptyenv}{\emptyenv}{\termone}}}
  \Big)} 
  \ar[d] & \mathtt{F}({\lambdatraceleq}) & 
  \effunit{\Big(
  \pairconf{\ctxone}{\effunit{\econfterm{\emptyenv}{\emptyenv}{\termtwo}}}
  \Big)} \ar[d]
  \\
  \eff{\gop}{n}{i}{\Big(
  \pairconf{\ctxtwo_i}{\effunit{\econfterm{\emptyenv}{\emptyenv}{\termone}}}
  \Big)}
  & ? 
  & \eff{\gop}{n}{i}{\Big(
  \pairconf{\ctxtwo_i}{\effunit{\econfterm{\emptyenv}{\emptyenv}{\termtwo}}}
  \Big)}
  }
  $$
  Next, we have the interaction between $\termone$ (resp. $\termtwo$)
  and the new contexts $\ctxtwo_i$s: 
  $$
  \xymatrix{
  \effunit{\Big(
  \pairconf{\ctxone}{\effunit{\econfterm{\emptyenv}{\emptyenv}{\termone}}}
  \Big)} 
  \ar[d] & \mathtt{F}({\lambdatraceleq}) & 
  \effunit{\Big(
  \pairconf{\ctxone}{\effunit{\econfterm{\emptyenv}{\emptyenv}{\termtwo}}}
  \Big)} \ar[d]
  \\
  \eff{\gop}{n}{i}{\Big(
  \pairconf{\ctxtwo_i}{\effunit{\econfterm{\emptyenv}{\emptyenv}{\termone}}}
  \Big)} \ar[d]
  & ? 
  & \eff{\gop}{n}{i}{\Big(
  \pairconf{\ctxtwo_i}{\effunit{\econfterm{\emptyenv}{\emptyenv}{\termtwo}}}
  \Big)} \ar[d]
  \\
  \eff{\goptwo}{m}{j}{\Big(
  \pairconf{\ctxtwo_j'}{\eff{\gop_1}{n_1}{i}{\econfone_i}}
  \Big)}
  & ? 
  & \eff{\goptwo}{m}{j}{\Big(
  \pairconf{\ctxtwo_j'}{\eff{\gop_2}{n_2}{i}{\econftwo_i}}
  \Big)}
  }
  $$
  where $\eff{\gop_1}{n_1}{i}{\econfone_i}
  \kappastartraceleq
  \eff{\gop_2}{n_2}{i}{\econftwo_i}$, and thus 
  $\eff{\gop_1}{n_1}{i}{\econfone_i}
  \kappasimilarity
  \eff{\gop_2}{n_2}{i}{\econftwo_i}$.
  At this point the invariant should be clear. All $\strongkleisli{\sysf}$-states in the 
  diagram have the same `outermost effect', argumentwise equal contexts, and 
  $\strongkleisli{\syskappa}$-similar
  inner ($\strongkleisli{\syskappa}$-)states. This directly leads us to the following 
  definition.
}

We thus have two $\strongkleisli{\sysf}$-states, $\mfstarone_{\termone}$ 
and $\mfstarone_{\termtwo}$, which are related by the obvious lifting of
$\lambdatraceeq$ to $\strongkleisli{\sysf}$-states. Such a lifting, however, 
is not preserved by the dynamics of $\strongkleisli{\sysf}$. 
In order to conclude the wished thesis, we need a stronger relation. 
In fact, it is sufficient to find a relation 
$\relone$ on $\strongkleisli{\sysf}$-states
such that 
\begin{itemize}
\item[(i)] $\mfstarone_{\termone} \relone \mfstarone_{\termtwo}$, 
\item[(ii)] $\relone$ is closed under the dynamics of $\strongkleisli{\sysf}$ 
(meaning that $\mfstarone \relone \mfstartwo$ implies 
$\semfstar{\mfstarone} \relone \semfstar{\mfstartwo}$), and 
\item[(iii)] $\relone$ respects $\obsfstar$, meaning that
$\mfstarone \relone \mfstartwo$ implies 
$\obsfstar(\mfstarone) = \obsfstar(\mfstartwo)$.
\end{itemize}
Actually, since $\obsfstar$ is continuous, we can
replace implication (ii) with 
\begin{itemize}
\item[(ii')] $\mfstarone \relone \mfstartwo$ implies 
$\forall k \geq 0.\ \semfstarn{\mfstarone}{k} \relone \semfstarn{\mfstartwo}{k}$. 
\end{itemize}
  Indeed, if that is the case, then by (iii) we have 
  $\obsfstar\ \semfstarn{\mfstarone}{k} = \obsfstar\ \semfstarn{\mfstartwo}{k}$, for 
  any $k \geq 0$ 
  We conclude:
  $$
  \obsfstar\ \semfstar{\mfstarone}
  =
  \obsfstar\ \lub_k \semfstarn{\mfstarone}{k}
  = 
  \lub_k \obsfstar\ \semfstarn{\mfstarone}{k} 
  = 
  \lub_k \obsfstar\ \semfstarn{\mfstartwo}{k}
  =
  \obsfstar\ \semfstar{\mfstartwo}
  $$

  Coming back to the task of finding the desired $\relone$, 
  we see that the latter is essentially given by the following diagram, 
  which is obtained by very definition of $\semfstar{-}$. 

  $$
  \xymatrix@C=0.5pc{
  \effunit{
  \pairconf{\ctxone}{\effunit{\econfterm{\emptyenv}{\emptyenv}{\termone}}}
  } 
  \ar[d] & \relone & 
  \effunit{
  \pairconf{\ctxone}{\effunit{\econfterm{\emptyenv}{\emptyenv}{\termtwo}}}
  } \ar[d]
  \\
  \eff{\gop}{n}{i}{
  \pairconf{t_i}{\effunit{\econfterm{\emptyenv}{\emptyenv}{\termone}}}
  } \ar[d]
  & \relone 
  & \eff{\gop}{n}{i}{
  \pairconf{t_i}{\effunit{\econfterm{\emptyenv}{\emptyenv}{\termtwo}}}
  } \ar[d]
  \\
  \eff{\goptwo}{m}{j}{
  \pairconf{s_j}{\eff{\Xi_j}{l_j}{a}{\econfone_a}}
  }
  & \relone
  & \eff{\goptwo}{m}{j}{
  \pairconf{s_j}{\eff{\Upsilon_j}{h_j}{b}{\econftwo_b}}}
  }
  $$

  Suppose we begin by evaluating the context $\ctxone$ in isolation, i.e. 
  without interacting with $\termone$ (resp. $\termtwo$) 
  This is nothing but the application of the first five rules in Figure~\ref{fig:operational-semantics-lambda-bang-open}, and corresponds 
  to moving from 
  the first to second raw in the above diagram. Notice that now states 
  consist of equal outermost (generic) effects and argumentwise equal open terms.

  Next, we have the interaction between $\termone$ (resp. $\termtwo$)
  and the new contexts $t_i$. 
  By inspection of the rules in Figure~\ref{fig:operational-semantics-lambda-bang-open},
  we see that $\termone$ and $\termtwo$, as well as the monadic configurations 
  coming from their evaluation, do not affect the outermost generic effects, 
  which, instead, are modified by the terms $t_i$ only, thus remaining equals. 
  Additionally, the open terms $t_i$ can be modified by 
  $\termone$ and $\termtwo$ (and their monadic configurations 
  coming from their evaluation) only by renaming variables and by replacing 
  stuck terms with variables. Such modifications correspond to 
  $\strongkleisli{\syskappa}$-actions only, and $\termone$ and $\termtwo$ having 
  the same traces, they can make the same modifications. We thus see that what we have 
  reached are $\strongkleisli{\sysf}$-states with the same outermost generic effect, 
  argumentwise equal open terms, and, 
  thanks to Proposition~\ref{prop:similarity-coincides-with-traceleq-bang},
   $\kappastartraceeq$-related inner $\syskappa$-states 
  (meaning that $\effj{\Xi_j}{l_j}{a}{\econfone_a}
  \kappastartraceeq
  \effk{\Upsilon_j}{h_j}{b}{\econftwo_b}$ in the diagram above). 
  This is our invariant.

By inspecting rules in Figure~\ref{fig:operational-semantics-lambda-bang-open}, 
we see that the desired relation is thus defined as follows:
\begin{definition}
Let $\relone$ be a $\strongkleisli{\syskappa}$-relation. Define the 
$\strongkleisli{\sysf}$-relation $\bc{\relone}$, called the
\emph{Barr and contextual closure} of $\relone$, as:
$$
\bc{\relone} \defeq 
\big \{\big ( \eff{\gop}{n}{i}{t_i;\mconfone_i},
\eff{\gop}{n}{i}{t_i;\mconftwo_i} \big )
\mid \gop \in \monad(\makeset{n}) \wedge \forall i.\ \mconfone_i \relone \mconftwo_i \big\}.
$$
\end{definition}

\shortv{
  Relying on Proposition~\ref{prop:similarity-coincides-with-traceleq-bang},
  we can prove (by induction on $k$) our main lemma.
}

\begin{lemma}[Main Lemma]
\label{lemma:main-lemma}
For all $\strongkleisli{\sysf}$-states $\mfstarone, \mfstartwo$:
$$\mfstarone \bc{\kappastartraceeq} \mfstartwo \implies 
\forall k \geq 0.\ \semfstarn{\mfstarone}{k} 
\bc{\kappastartraceeq} \semfstarn{\mfstartwo}{k}.
$$
\end{lemma}

  The proof of Lemma~\ref{lemma:main-lemma} follows the informal 
  intuition given in the above discussion, and proceeds by induction on 
  $k$ taking advantage of the equality ${\kappastartraceeq} = 
  {\kappastarbisimilarity}$.

Finally, since $\termone \lambdatraceeq \termtwo$ obviously implies 
$\mfstarone_{\termone} \bc{\relone} \mfstarone_{\termtwo}$,
we obtain soundness of $\lambdatraceeq$ for contextual equivalence.

\begin{theorem}
\label{thm:soundness}
${\lambdatraceeq} \subseteq {\ctxeq}$.
\end{theorem}

As already anticipated, ${\lambdatraceeq}$ is also complete 
for ${\ctxeq}$. This is proved by showing that ${\ctxeq} \subseteq {\lambdatraceeq}$, 
which is itself proved by noticing that any $\syskappa$-action can be encoded as 
a context. 
\begin{theorem}
\label{thm:full-abstraction}
  ${\ctxeq} = {\lambdatraceeq}$.
\end{theorem}

We omit the proof of this result, as the encoding of $\syskappa$-actions 
as contexts is essentially the same one
of \cite{CrubilleDalLago/ESOP/2017}.
}

\section{Conclusion and Future Work}

In this paper, we have introduced resource transition systems 
as an operational account of both intensional and 
extensional behaviours of linear effectful programs 
with explicit copying. On top of resource transition systems, 
we have defined trace equivalence and showed that the latter 
is fully abstract for contextual equivalence. 

Although the present paper focuses on linearity (and effects), 
the authors' believe that resource transition systems can be 
extended to deal with finer notions of context dependence 
such as \emph{structural coeffects} 
\cite{Mycroft-et-al/ICFP/2014,Gaboradi-et-al/ICFP/2016,Brunel-et-al/ESOP/2014,Orchard:2019:QPR:3352468.3341714}. To do so, one should modify resource 
transition systems by considering sequences 
of terms indexed by elements of 
a resource algebra (the latter being a preordered semiring),
and let transitions update resources. 
Thus, for instance, from a 
sequence $(\gone, \lan\termone\ran_{r+1}, \done)$, 
meaning that $\termone$ is available according to the resource $r+1$, we have a transition to
$\econf{\gone,\lan\termone\ran_{r},\done}{\termone}$. 

The authors also believe that resource transition systems can be used 
to generalise Crubill\'e and Dal Lago probabilistic program metric  
to arbitrary algebraic effects. To do so, one would simply replace 
ordinary relations with relations taking values over quantales
\cite{Gavazzo/LICS/2018}.

Finally, as a long term future work, the authors would like to 
whether the ideas presented in this paper 
can be adapted to deal with quantum languages \cite{DBLP:journals/mscs/SelingerV06,DBLP:conf/fossacs/SelingerV08}, where the interaction between linearity and effects plays a 
central role. In fact, although we have not discussed tensor product types (which 
play a crucial role in a quantum setting), it is not hard to see that 
resource transition systems can be extended to deal with such types \cite{CrubilleDalLago/LICS/2015}.

\subsection{Related Work}

This is not the first work on operationally-based 
notions of program equivalence for linear calculi. 
In particular, notions of equivalences have been defined by means of 
logical relations by Bierman, Pitts, and Russo 
\cite{DBLP:journals/entcs/BiermanPR00}, of applicative bisimilarity 
by Bierman \cite{DBLP:journals/jfp/Bierman00} and Crole\footnote{Crole's applicative 
bisimilarity, however, does not deal with copying.} \cite{DBLP:journals/igpl/Crole01}, 
of trace equivalence by Deng and Zhang \cite{DBLP:journals/tcs/DengZ15,DBLP:conf/tase/DengF17}, as well as of a number of possible worlds-indexed equivalences 
(e.g. \cite{DBLP:journals/fuin/AhmedFM07,DBLP:conf/popl/KrishnaswamiPB15}). 
As already remarked, one of the advantages of resource transition systems 
(and their associated trace equivalence) compared, e.g., with logical relations,
is that they they provide a \emph{first-order} account of program equality. 

Among first-order notion of program equivalence, 
Bierman's applicative bisimilarity plays a prominent role. 
The latter is a lightweight extensional equivalence
extending Abramsky's applicative bisimilarity \cite{Abramsky/RTFP/1990}
to a \emph{pure} linear $\lambda$-calculus with explicit copying 
Bierman's applicative bisimilarity can be readily extended to calculi 
with algebraic effects along the lines of \cite{DalLagoGavazzoLevy/LICS/2017}, 
this way obtaining a notion of equivalence invalidating \eqref{bang-dist}.
However, such a notion of bisimilarity
stipulates that two programs $\bang{\termone}$ and $\bang{\termtwo}$ 
are bisimilar if and only if $\termone$ and $\termtwo$ are, this way 
making bisimilarity insensitive to linearity, and thus invalidating
\eqref{lambda-dist} as well\footnote{Besides, notice that bisimilarity being sensitive 
to branching, it naturally invalidates \eqref{lambda-dist}.}. 

Deng and Zhang's linear trace equivalence has been designed to 
study the interaction of linearity and (both pure and probabilistic) 
nondeterminism. The latter equivalence, in fact, validates 
\eqref{lambda-dist}. However, linear trace equivalence does 
not deal with (explicit) copying: even worst, natural extensions of such notions 
to languages with copying result in equivalences validating 
\eqref{bang-dist}. 
Crubill\'e and Dal Lago  \cite{CrubilleDalLago/ESOP/2017} 
solved that problem by introducing a
tuple-based applicative bisimilarity for a calculus with probabilistic 
nondeterminism and explicit copying. Our notion of a resource transition 
system can be seen as a generalisation of the Markov chain underlying 
tuple based applicative bisimilarity to arbitrary algebraic effects. 

\bibliographystyle{plain}
\bibliography{main}

\end{document}